\documentclass{aa}

\usepackage{natbib}
\usepackage{colortbl}
\definecolor{Gray}{gray}{0.9}
\usepackage{graphicx}
\usepackage{float}
\usepackage{txfonts}
\usepackage{hyperref} \hypersetup{ colorlinks=true, citecolor={blue} ,linkcolor=blue, filecolor=blue, urlcolor=blue, pdftitle={Overleaf Example}}

\newcommand{\bg}[0]{Br\textrm{$\gamma$}}

\usepackage{xcolor}
\usepackage{ulem}

\definecolor{mygray}{gray}{0.9}

\begin{document} 



  \title{The interplay between disk wind and magnetospheric accretion mechanisms in the innermost environment of RU Lup}




   \author{J. A. Wojtczak \inst{1}
        \and B. Tessore\inst{2}
        \and    L. Labadie\inst{1}
        \and    K. Perraut\inst{2}   
        \and    J. Bouvier\inst{2}
        \and    C. Dougados\inst{2}
        \and    H. Nowacki\inst{2}
        \and    A. Soulain\inst{2}
        \and    E. Alécian\inst{2}
        \and    G. Pantolmos\inst{2}
        \and    J. Ferreira\inst{2}
        \and    C. Straubmeier\inst{1}
        \and    A. Eckart\inst{1}
        }

\institute{I. Physikalisches Institut, Universität zu Köln, Zülpicher Str. 77, 50937, Köln, Germany \\
\email{wojtczak@ph1.uni-koeln.de}
\and
Univ. Grenoble Alpes, CNRS, IPAG, 38000 Grenoble, France 
} 
   \date{Received -; accepted -}

 
  \abstract
  {Hydrogen recombination lines such as \bg \  are tracers of hot gas within the inner circumstellar disk of young stellar objects (YSOs). In the relatively cool innermost environment of T Tauri stars specifically, \bg \ emission is closely associated with magnetically driven processes, such as magnetospheric accretion. Magnetospheric emission alone would arise from a relatively compact region that is located close to the co-rotation radius of the star-disk system. Since it was previously found that the \bg \ emission region in these objects can be significantly more extended than this, it was speculated that \bg \ emission may also originate from a larger structure, such as a magnetised disk wind.}
  { Our aim is to build upon the analysis presented in our previous work by attempting to match the observational data obtained with VLTI GRAVITY for RU Lup in 2021 with an expanded model. Specifically, we will determine if the inclusion of an additional disk wind as a \bg \ emitter in the inner disk will be able to reproduce the trend of increasing sizes at higher velocities. In addition, we will investigate whether the additional component will alter the obtained photocentre shift profiles to be more consistent with the observational results.}
  {We make use of the MCFOST radiative transfer code to solve for \bg \ line formation in the innermost disk of an RU Lup-like system. From the resulting images we compute synthetic interferometric observables in the form of the continuum-normalised line profiles, visibilities, and differential phases. Based on these computations, we first investigate how individual parameter variations in a pure magnetospheric accretion model and a pure parameteric disk wind model translate to changes in these derived quantities. Then we attempt to reproduce the RU Lup GRAVITY data with different parameter variants of magnetospheric accretion models, disk wind models, and combined hybrid models.}
  {We  demonstrate that magnetospheric accretion models and disk wind models on their own can emulate certain individual characteristics from the observational results, but individually fail to comprehensively reproduce the observational trends. Disk wind plus accretion hybrid models are in principle capable of explaining the variation in characteristic radii across the line and the corresponding flux ratios. While the model parameters of the hybrid models are mostly in good agreement with the known attributes of RU Lup, we find that our best-fitting models deviate in terms of rotational period and the size of the magnetosphere. The best-fitting hybrid model does not respect the co-rotation criterion, as the magnetospheric truncation radius is about 50\% larger than the co-rotation radius. }
  {The deviation of the found magnetospheric size when assuming stable accretion with funnel flows indicates that the accretion process in RU Lup is more complex than what the analytical model of magnetospheric accretion suggests. The result implies that RU Lup could exist in a weak propeller regime of accretion, featuring ejection at the magnetospheric boundary. Alternatively, the omission of a large scale halo component from the treatment of the observational data may have lead to a significant overestimation of the emission region size.}

   \keywords{stars: formation - stars: variables: T~Tauri - accretion, accretion disks -  techniques: high angular resolution - techniques: interferometric - }
   
 \authorrunning {J.~A.~Wojtczak et al.}
    \titlerunning{Disk wind and magnetospheric accretion mechanisms in the innermost environment of RU Lup}
   \maketitle

\section{Introduction}

Top-tier optical/infrared interferometers, such as the Center for High Angular Resolution Astronomy (CHARA) Array or the Very Large Telescope Interferometer (VLTI), have benefitted from the great technical advancements in long baseline interferometry over the past 20 years to the point where once inaccessible regions close to the surface of a in young stellar object (YSO) can be explored in increasing detail.
The current generation of instruments now routinely allows astronomers to probe even the innermost disk regions of relatively faint sources, such as T Tauri stars, at angular scales in the sub-milliarcsond regime. Their improved capabilities present the astrophysical community with opportunities to put long held beliefs about the physical processes that govern the star-disk interface to the test. The past years have seen a growing number of studies focussing on spatially resolved observations of the innermost circumstellar disk YSOs \citep{Soulain2023,Ganci2021,Perraut2021, GarciaLopez2020,Setterholm2018}. Theoretical frameworks describing the dynamics of ejection and accretion on such small scales have long been discussed in the context of spectroscopic surveys (e.g. \cite{Muzerolle1998, Muzerolle2001}), but only now does long baseline interferometry provide the means to also directly trace the spatial signatures of those mechanisms. \\ \indent
The investigation of the magnetospheric accretion paradigm, according to which the flow of matter close to magnetically active YSOs should be heavily dominated by the stellar magnetic field, has been an important starting point in this context. For T Tauri stars with their kG order magnetic field strengths in particular, the notion is that matter would be funnelled from the inner edge of a magnetically truncated disk along the magnetic field lines onto the stellar surface at high latitudes, see the detailed treatments in, for example, \cite{Bouvier2007} and \cite{Hartmann2016}. The temperatures and densities within these funnel flows would then give rise to higher order hydrogen line emission, as in the form of the \bg \ line, which would otherwise be absent in the relatively cool environment of classical T Tauri stars. \\ \indent
Initial observations of T Tauri YSOs such as TW Hya \citep{GarciaLopez2020} and DoAr44 \citep{Bouvier2020a} with VLTI GRAVITY confirm the assumption that \bg \ emission in these systems could thus be spatially constrained to a condensed region close to the stellar surface. 
In both cases, the authors relied on the co-rotation criterion to determine whether the spatial scale of the emission region is consistent with the scenario of stable magnetospheric accretion. The criterion dictates that the magnetosphere needs to be truncated within the co-rotation radius $r_{co}$ for stable accretion columns to form, i.e. $r_{mag} \leq r_{co}$ \citep{Romanova2016}.
In \cite{Wojtczak2023} we follow up on these earlier works with a comparative study of \bg \ line emission in seven T Tauri stars with strong \bg \ signals. There the interpretation of the derived interferometric sizes and positional offsets of the \bg \ emission region is supplemented by the use of a simple axisymmetric magnetospheric accretion model. This model is used to produce synthetic interferometric observables in order to investigate and identify the characteristic behaviour of the magnetospheric accretion paradigm as seen through the lens of interferometry.
As one of the key results, the study finds that, while the weakest accretors in the sample, TW Hya and DoAr 44, are in good agreement with the accretion model and co-rotation criterion, the remaining targets show emission on spatial scales beyond typical magnetospheric radii.  \\ \indent
Given the apparent limitations of the axisymmetric magnetospheric accretion scenario to explain many of the spectro-interferometric signatures obtained with GRAVITY, it is obvious that a more complex model is needed to approximate the observations. Firstly, the assumption of axisymmetry in the model could be questioned. It is well established that the magnetic dipole in many magnetically active YSOs is tilted with respect to the stellar rotational axis \citep{Donati2007, Johnstone2014,McGinnis2020}, meaning real observed systems typically exhibit non-axisymmetric magnetospheric geometries. However, while a non-zero dipole tilt may very well have some degree of impact on the interferometric signatures, it is not clear how this in itself would resolve the discrepancy between the extended emission region size and the comparatively small co-rotation radius. It is more likely that an additional large scale \bg \ emission component, such as a disk wind being launched from the magnetosphere-disk interface, is needed to address this issue.  \\ \indent
Such winds, among other outflows, have been thoroughly discussed in the context of young stars for decades due to their connection to large scale jet structures that are observed observed around multiple YSO systems, such as RU Lup or AS 353 \citep{Takami2003, Whelan2021}, as well as their presumed role in managing excess angular momentum in star formation \citep{HartmannStauffer1989,Shu1994,Ferreira2000,MattPudritz2005}. Winds constitute a strong contributor to mass loss and thus disk dispersal in YSOs as these objects evolve towards the main sequence \citep{Alexander2014,Tabone22}. As such, accretion and ejection are fundamentally related and mass ejection rates are typically proportional to the accretion rate. Both are known to decline with the age of the YSO as the system matures out of its pre-main sequence (PMS) phase (see e.g. \cite{Watson2016}). In \cite{Wojtczak2023}, we show that the younger systems from among the sample, with higher accretion rates, deviate from the pure magnetospheric accretion scenario more strongly. The idea of a wind component potentially acting as an additional emitter of \bg \ radiation in younger systems, then diminishing in relative strength with increasing age, could serve as a possible explanation of the distinction we find among the sample objects. \\ \indent
Previous theoretical works by, for example, \cite{Garcia2001} and \cite{Lima2010} suggest that a magnetically driven wind, launched through magnetohydrodynamical (MHD) processes, could reach temperatures on the order of $10^4 \ K$ close to the disk surface. Such an MHD wind could thus provide a sufficiently hot and dense environment to produce a significant amount of \bg \ radiation on spatial scales close to the magnetosphere but larger than the co-rotation radius \citep{Romanova2015}.
MHD wind models have in the past been tested predominantly against spectroscopic observational data \citep{Wilson2022, Weber2020}, whereas interferometric studies that incorporate radiative transfer simulations are so far scarce. Works that feature models of wind or accretion based hydrogen emission are largely restricted to Herbig Ae/Be stars (e.g. \cite{Weigelt2011,Kurosawa2016}), although more recent contributions have begun to focus on connecting simulations and interferometry in the context of classical T Tauri stars \citep{Tessore23, Wojtczak2023}. \\ \indent
In this paper, we build not only on our own previous work, but expand upon the body of interferometric studies of inflow/outflow mechanisms in general (e.g. \cite{Eisner2010}). For the first time, we combine sophisticated modelling efforts of the near-infrared (NIR) signatures of disk winds and magnetospheric accretion in T Tauri stars with the unparalleled spatial resolution of long-baseline interferometry.  We re-investigate the comparison of synthetic spectroscopic and interferometric observables to VLTI GRAVITY observational data for the exemplary case of the young star RU Lup in the light of this comprehensive modelling approach. The goal of the analysis is to determine whether these types of complex wind-magnetospheric accretion hybrid models are suited to address the 
discrepancies between the simple accretion scenario and the RU Lup data.  \\ \indent
Section 2 of this work recapitulates the most important observational results on RU Lup.
Section 3 details the use of our models. This includes an introduction to the MCFOST radiative transfer code and the technical choices made in preparation of the simulations, as well as a description the model parameters and of the treatment of the synthetic interferometric data. 
Section 4 contains the results of our simulations in the form of six exemplary model configurations and their comparison to the RU Lup GRAVITY data.
Section 5 gives an overview of the effects of parameter variations in the context of the synthetic interferometric data in more general terms.
Section 6 presents a thorough discussion of the most important results.




\section{Observational results on RU Lup}
\begin{figure}[h] 
    \centering
    \includegraphics[width=0.8\linewidth]{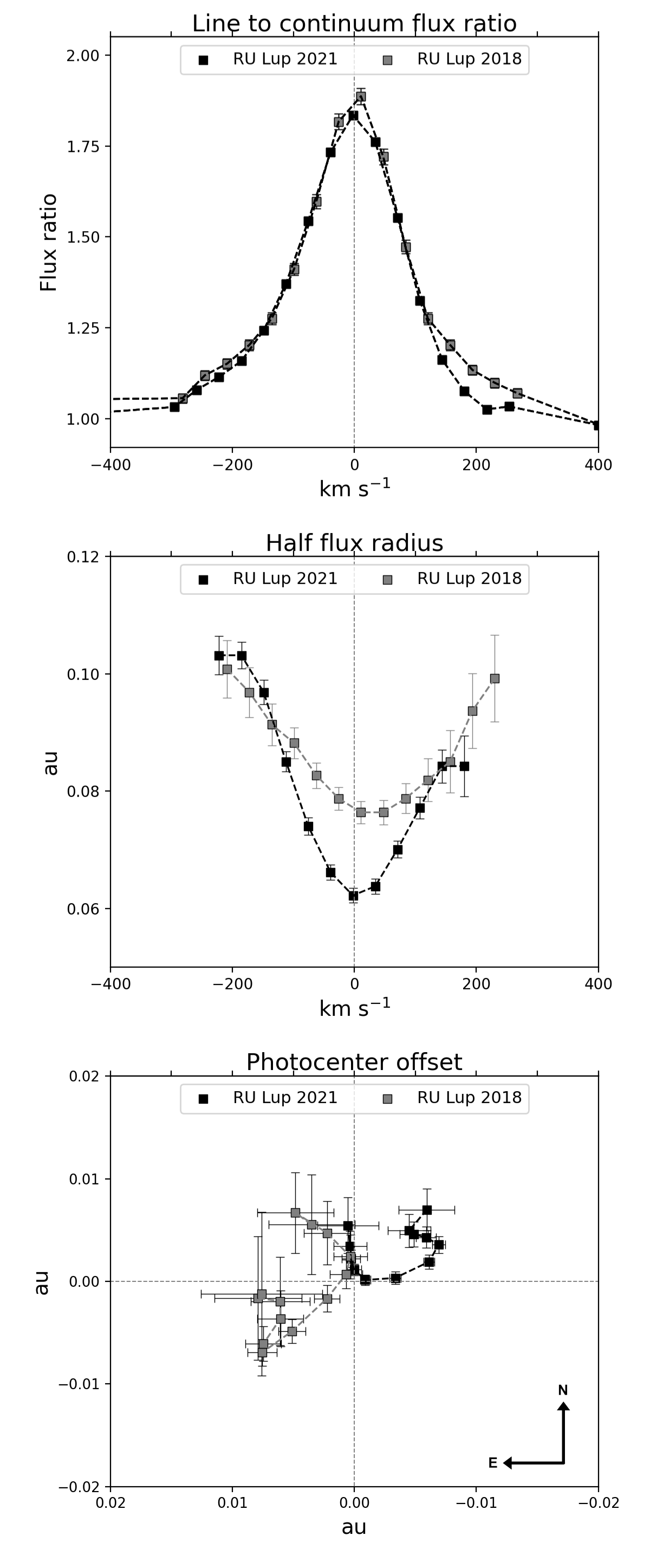}
    \caption{Previous results for CTTS RU Lup, derived from VLTI GRAVITY observations in 2018 and 2021. Shown here are (from top to bottom) the line-to-continuum flux ratio of the \bg \ spectral line, the characteristic size of the emission region across the velocity channels of the line, and the corresponding positions of the emission region photocentres. 
    }\label{fig:dataplot}
  \end{figure}
  \noindent
\label{sec:Obs}

RU Lup is a classical T Tauri star, located in the Lupus star forming region at a distance of 157.5 pc \citep{GaiaDR3}. The system is observed at an almost pole-on configuration with an inclination of only 18.8$^{\circ}$ according to an interferometric study of the large-scale disk structure in the 1.25 mm continuum with ALMA \citep{Huang2018}.
RU Lup has a known large scale outflow component in the form of a jet whose blue-shifted component has been determined to be at a position angle of 229$^{\circ}$ north to east \citep{Whelan2021}. It is generally considered to be a strong accretor, with mass accretion rates in the literature ranging up to $30*10^{-8} \ M_{\odot} yr^{-1}$  \citep{Siwak2016}. \cite{Stock22} proposed that the accretion rate of RU Lup, as estimated from an empirical relationship between line luminosity and accretion luminosity for certain tracing spectral lines, can vary by as much as a factor of two over a 15 day period, making accretion in RU Lup potentially also highly variable. (Sousa et al. 2024, in prep) \\ \indent
RU Lup was observed with GRAVITY in April 2018 and May 2021, combining the four 8.2 m Unit Telescopes (UTs) of the VLTI to record spectrally dispersed interferometric fringes with 6 baselines. The observations made use of GRAVITY's single field mode, delivering simultaneously low spectral resolution (R $\sim$ 22) fringe tracker (FT) and high spectral resolution (R $\sim$ 4000) science channel (SC) data of RU Lup across the NIR K-band between 1.98 $\mu m$ and 2.4 $\mu m$. The effective baseline lengths of the observation range from 46.48 m (UT3-UT2) up to 128.5 m (UT4-UT1), translating into maximum angular resolutions in the K-band between around 4.6 and 1.7 mas, since the angular resolution scales with the projected baseline length B as $\frac{\lambda}{2B}$. In both 2018 and 2021, the total observation time ran for a little over an hour, preceded by a single calibrator measurement in each case. \\ \indent
In the following we summarise the most important results from the GRAVITY continuum and \bg \ line studies on RU Lup \citep{Perraut2021,Wojtczak2023}.
The continuum visibility data for RU Lup was fitted with a multi-component geometrical model \citep{Berger2007}. It consists of a point source to account for the star itself, a ring-like region which represents the interferometrically partially resolved region around the inner dusty rim, and an extended component which we refer to as the 'halo'. The halo component is thought to represent the large scale scattered light emission from the disk. It takes into account K-band continuum flux contributions which appear in the spectrum, but are spatially overresolved at the projected baseline lengths of the telescope configuration. The relevance of the halo and the continuum disk will be further detailed in Section \ref{sec:interfero}. The ring-like morphology of the compact environmental component was chosen due to the association of the K-band continuum emission with the hot dusty wall at the sublimation radius. The continuum analysis yielded a halo contribution of 12 $\pm$ 3 \%  and a ring contribution of 30 $\pm$ 10 \% to the total continuum emission. The fit of the ring region, for which a Gaussian brightness profile was assumed, yielded a size in the form of a half width at half maximum (HWHM), or half flux radius, of 0.21 $\pm$ 0.06 au in 2018 and 2021. The inclination of the inner disk continuum was determined to be 16$^{+6}_{-8}$ degrees for the 2018 data and 20$^{+6}_{-8}$ degrees for 2021 data. The position angles of the ring were constrained at 99 $\pm$ 31 degrees and 101 $\pm$ 31 degrees for 2018 and 2021, respectively. \\ \indent
Figure \ref{fig:dataplot} shows the observational results obtained for the \bg \ emission region.
The \bg \ line profile shows a small variation in its equivalent width (-16.82 $\AA$ to -13.47 $\AA$) from 2018 to 2021, features an asymmetric shape with blueshifted excess emission, and no visible inverse P Cygni characteristic.
The visibility data was fitted using a geometrical Gaussian disk model. As such, we presumed that the brightness distribution of the emission region could be described by a disk-like morphology with a centrally peaked brightness profile, which then falls off as a Gaussian function with increasing distance from the centre. In this manner we obtained spectrally dispersed characteristic sizes, again in the form of a half flux radius, for up to 13 velocity channels across the emission line. Similarly, we derived the positional offset of the barycentre of the brightness distribution (i.e. the photocentre) for the same spectral channels from the continuum-subtracted differential phase data. Both the visibility and differential phase data were corrected for continuum contributions, meaning the resulting quantities describe the pure line \bg \ emission region. Hereby we found that the \bg \ half flux radii (2018: 6.32 R$_*$, 2021: 5.01 R$_*$ in the centre channel) were more extended than the co-rotation radius (R$_{co}$ = 3.31 R$_*$ for P$_*$=3.7 d, R$_*$=2.6 R$_{\odot}$, and M=0.6 M$_{\odot}$, as assumed in \cite{Perraut2021} and \cite{Wojtczak2023}), by a factor of more than 1.5. In addition, the emission region sizes were increasing  towards the high velocity channels, which was inconsistent with the centrally peaked size profile predicted by the simple axisymmetric accretion model. The distribution of emission region photocentres across the line showed no clear alignment with the known jet axis or the model profile. The magnitude of the displacements was significantly smaller, by a factor of about 4 in the most extreme channel, than that obtained from the synthetic model data.   \\ \indent
These clear deviations from the simple magnetospheric accretion scenario make RU Lup a suitable test case to investigate the effect of an additional \bg \ emitting disk wind on the synthetic observables. The low inclination configuration, confirmed by GRAVITY, promises to be particularly advantageous compared to some of the other objects from the GRAVITY T Tauri sample for which we found similar results, given that it more clearly separates the accretion and ejection emission regions when the system is viewed near pole-on. In this work we focus on the comparison of the new radiative transfer simulations to the 2021 dataset, since the fundamental trends in the observable profiles are retained between 2018 and 2021. In both cases, we detect an asymmetric line, a spatially extended emission region with increasing sizes at high velocities and a compact multi-axial distribution of emission region photocentres across the \bg \ line.
For a more detailed description of the reduction process, as well as a more thorough discussion on the results for both continuum and line emission data, we refer to \cite{Perraut2021,Wojtczak2023}.

\section{Synthetic data modelling}

In the following we introduce the methodology used to compute synthetic interferometric observables from the analytical magnetospheric accretion and disk wind models we used. We first introduce the radiative transfer code MCFOST, before we give a succinct description of the models and their relevant parameters. This section concludes then with a description of the model observable computation and treatment.
\subsection{Radiative transfer code}
\begin{figure}[t!] 
    \centering
    \includegraphics[width=1\linewidth]{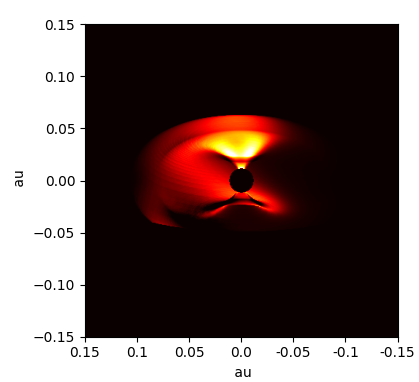}
    \caption{Example of a magnetospheric accretion model image, produced with the MCFOST radiative transfer code. Shown here is the continuum-subtracted \bg \ emission region of a purely axisymmetric rotating magnetosphere with no additional emission components. Channel line maps such as this can be used to compute artificial interferometric observables which we are then able to compare to real observational data. The image here specifically shows the line map for the -69 km/s velocity channel.
    }
    \label{fig:Mag70Deg}
  \end{figure}
  \noindent
We made use of MCFOST \citep{Pinte2006,Pinte2009} to produce intensity maps, such as shown in Fig. \ref{fig:Mag70Deg}, in the wavelength range of  the \bg \ line. MCFOST is a dust and atomic line Monte Carlo and ray-tracing based radiative transfer code. It is capable of computing both line and continuum fluxes in multi-level atomic systems under non-local thermodynamical equlibrium (non-LTE) conditions \citep{Tessore21,Tessore23}. The atomic hydrogen model used here includes the ground state and 15 excited bound states, leading to a total of 101 discrete bound-bound transitions and 15 bound-free transitions, of which the latter contribute to the total system continuum emission while the former produce discrete spectral line emission. While this hydrogen model file contains a large number of other prominent hydrogen lines, such as H$\alpha$, we only consider line emission produced by the \bg \ electronic 7-4 transition at a vacuum wavelength of 2.16612 $\mu m$. \\ \indent
The images were computed across 101 wavelengths, centred on the line position, with a channel width of 0.723 $\AA$. We chose to compute the atomic populations and radiative transfer on a spherical grid with a logarithmic distribution of points in the radial direction to maximise the density of cells in the central regions of the image, where the relatively complex brightness distribution of the star-magnetosphere system is located. As the axisymmetric magnetospheric model and the wind model are also symmetric with respect to rotation about the stellar axis, they were computed on a 2D spherical grid in order to save on computation time. By contrast, the non-axisymmetric magnetospheric accretion models required the use of a 3D spherical grid due to the azimuth-dependence of the hemispheric accretion flows under a tilted magnetic dipole. On the 2D grids we set the number of radial and polar angle grid points to 150 each, while on the 3D grids an additional 64 points in azimuth are included. We confirm, based on comparative simulations of an axisymmetric model
on both a 2D and 3D grid, that resolution effects due to the smaller number of azimuth points do not affect the results. The extent of the grid constitutes a cut-off for the contributions coming from the outer regions of the wind, so the grid needs to be sufficiently large to include all significant flux, but should not be larger than necessary for the sake of computational time efficiency. To ensure this, the outer edge of the grid was tied to a multiple of the outer radius of the model. The exact number depends on the model used and was chosen by manually increasing the grid size until further increases did not significantly change the resulting line profile any longer. The effective range of grid sizes for the models discussed in section 4 lies between 36 and 47 R$_*$.
The final images were produced for multiple inclinations and, where applicable, azimuth angles. The field of view of the image was also anchored to the outer radius of the model. It was again set to be sufficiently wide to capture the entire flux at any possible inclination, but as small enough to minimise the resulting file sizes. The pixel size was, due to similar considerations as for grid size and field of view, set to a constant 0.03 au per pixel, meaning the number of pixels per image varies with the radial size of the chosen model. \\ \indent
Combining two different model components, such as the accretion region and the disk wind, necessitates a careful definition of both model regions via their geometric parameters as to avoid any kind of density overlap between them. Adding another model to the grid will overwrite existing cell information, meaning that any given cell can only be filled with material from either the accretion column or the wind. Beyond this point, all interactions between the different components come purely in the form of photon propagation along the line of sight through the different regions, by which way they may interact with each other. 
\begin{figure*}[t] 
    \centering
    \includegraphics[width=0.7\linewidth]{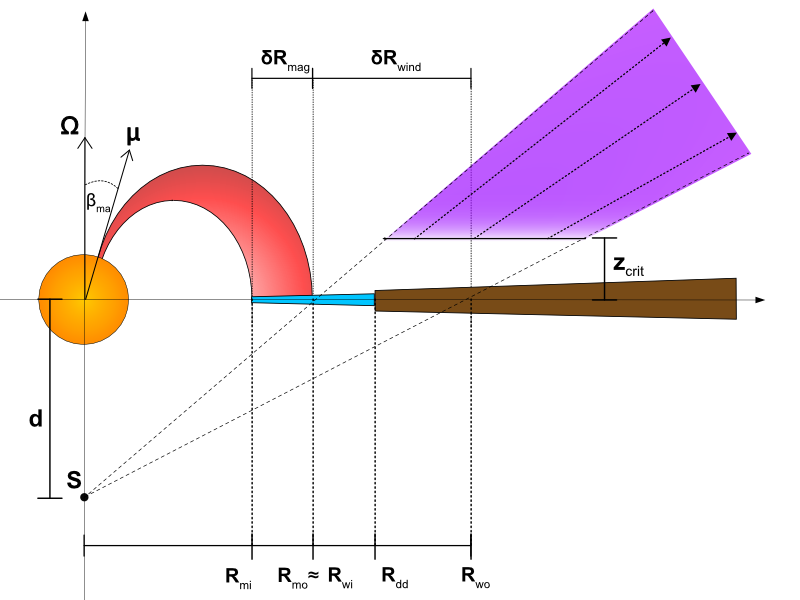}
    \caption{Schematic depiction of the magnetospheric accretion and disk wind models. Shown here are the geometrical parameters defining each component. These are: The rotational axis $\boldsymbol{\Omega}$, the magnetic dipole axis $\boldsymbol{\mu}$, the magnetic obliquity $\boldsymbol{\beta_{ma}}$, the wind focal point (\textbf{S}) displacement \textbf{d}, the inner and outer radius $\boldsymbol{R_{mi}}$ and $\boldsymbol{R_{mo}}$, and width  $\boldsymbol{\delta R_{mag}}$, of the magnetosphere, the inner and outer radius $\boldsymbol{R_{wi}}$ and $\boldsymbol{R_{wo}}$, and width $\boldsymbol{\delta R_{wind}}$, of the wind, the dark disk radius $\boldsymbol{R_{dd}}$ beyond which the disk becomes completely opaque, and the cutoff $\boldsymbol{z_{crit}}$ at which the isothermal wind instantly reaches its 10$^4$ K temperature. \\
    The lower hemisphere parts of wind and magnetospheric funnel are omitted in this depiction for the sake of clarity. \label{fig:WindMagSchem}
    }
\end{figure*}

\subsection{Magnetospheric accretion model} \label{sec:Macc}

Under the magnetospheric accretion paradigm, the inner disk is magnetically truncated at a certain distance from the central star, at which point gas is funneled along the magnetic field lines and transported onto the stellar surface near the poles. In order to compute the emission produced in such a system with the MCFOST code, we employ the model presented in  \cite{Hartmann1994,Kurosawa2006,Kurosawa2011}, and \cite{Lima2010} to define the hydrogen mass density $\rho$, the temperature profile T(R), and the poloidal velocity $v_p$ along the field lines. We refer to those works for a more quantitative description of the model. \\ \indent
In the axisymmetric scenario, the system is completely characterised by any 2D slice of the 3D distributions and can be parameterised on a 2D spherical coordinate grid centred on the stellar position. 
The fundamental assumption of magnetospheric accretion is that the stream of matter within the magnetosphere follows the geometry of the magnetic dipole field lines. 
The gas is assumed to have no velocity at the truncation radius and then accretes onto the star under the gravitational pull, which defines then the absolute velocity parallel to the field line rooted at the anchor point $R_m$. Since the magnetospheric funnel is not infinitely thin, but rather extends over a range of anchor points, we define an inner and an outer magnetospheric radius $R_{mi}$ and $R_{mo}$, respectively, and a width $\delta R$ so that $R_{mo}=R_{mi}+\delta R$. The velocity along any field line with an anchor point $R_m$ between these boundaries is fully defined by the free fall motion, depending on the stellar mass and radius. In addition, we assume that the magnetosphere rotates as a rigid body, so that the velocity component in the plane perpendicular to the field lines is determined by the stellar rotational period, leading to line broadening effects at non-zero inclinations.
The velocity field and funnel geometry also define a shock region on the stellar surface, where a certain amount of accreting material reaches the star per unit time and releases its energy.
The density along the accretion funnels is normalised to ensure consistency between this local mass flux onto the star and the global mass accretion rate $\dot{M}_{acc}$. 
The shock region itself is another source of continuum emission, acting as an additional black body with a temperature determined by the energy released by the infalling material. \\ \indent
With regards to the temperature profile in the funnel flows, we follow \cite{Kurosawa2006} and \cite{Lima2010} by adapting the \cite{Hartmann1994} temperature profile based on a volumetric heating rate $\propto \frac{1}{r^{-3}}$. 
The temperature along the funnel is then computed based on an assumed energy balance between the radiative cooling rates presented in \cite{Hartmann1982} and the volumetric heating rate. The profile is normalised to a maximum temperature along the funnel $T_{mag}$, which we set as a parameter in our model. \\ \indent
For the non-axisymmetric scenario depicted in Fig. \ref{fig:WindMagSchem}, we introduce an additional free parameter: the dipole tilt angle $\beta_{ma}$ relative to the direction of the stellar rotational axis. We refer to this angle as the 'magnetic obliquity' of the system. The dipole tilt also leads to a toroidal component of the magnetic field, which interacts with the velocity field in the plane perpendicular to the field lines by adding an additional toroidal velocity component. For a more thorough description of the density profile and magnetic field components in the non-axisymmetric case, we refer to \cite{Tessore23} and \cite{Mahdavi1998}.
In summary, our magnetospheric accretion model is fully characterised by the \textbf{stellar parameters} $R_*$, $M_*$, $P_*$, the \textbf{mass accretion rate} $\dot{M_{acc}}$, the \textbf{inner anchor point} $R_{mi}$ and \textbf{width} $\delta R_{mag}$ of the accretion columns, and the \textbf{obliquity} $\beta_{ma}$.

\subsection{Disk wind model} \label{sec:dwmodel}

In order to ensure a high degree of flexibility when adjusting the outflow component in the system, we chose to adapt the kinematic disk wind model described by \cite{Knigge1995}. Their model follows the basic principle of the magneto-centrifugal disk wind as proposed by \cite{BP1982}, which features a mass outflow arising from a disk in Keplerian rotation along a range of open field lines anchored at the disk midplane. It is fundamentally a parametric description of a disk wind, designed to allow for straightforward manipulation of the parameter space to be explored, whilst still approximately reproducing the attributes of a proper, self-consistently treated, MHD wind model. We refrain from reproducing the exact equations here, as they have been introduced in great detail not only in the original publication, but also other works such as \cite{Kurosawa2006} and \cite{Kurosawa2016}\\ \indent
The geometry of the model is biconical, featuring gas streams which follow a set of magnetic field lines lying on conically shaped surfaces, see Fig. \ref{fig:WindMagSchem}. This structure is well described by three geometric parameters, which come in the form of the focal point distance d, as well as an inner and outer wind boundary radius $R_{wi}$ and $R_{wo}$. The focal point (S in Fig. \ref{fig:WindMagSchem}) defines the origin of the magnetic field lines for one hemisphere, which is vertically displaced from the midplane along the disk axis by the focal point distance d. Adjusting this parameter allows us to effectively set the angle of the outflow and the degree of collimation between the wind stream lines. The inner and outer radii of the wind then give the radial distance from the star at which the closest and furthest field lines intersect the midplane, thus defining the effective wind region in our system as the set of conical surfaces between these two radii. \\ \indent
We further extend these parameters by an additional quantity $z_{crit}$, which describes the critical height z above the disk at which the wind reaches its final constant temperature of 10$^4$ K. It is depicted in Fig. \ref{fig:WindMagSchem} as the distance between the lower edge of the purple shaded wind region and midplane. Since our implementation treats the wind as isothermal, the critical height acts as a vertical lower end cutoff, effectively defining a starting height for the \bg \ emitting component. This approach, which essentially approximates a temperature profile along the vertical axis with little heating at first and then a very sharp rise in temperature close to $z_{crit}$, was chosen to simplify the question of the temperature structure in the gas streams. Below $z_{crit}$, the gas is assumed to be cold and effectively transparent at NIR K-band wavelengths.  The isothermal temperature of 10$^4$ K was chosen due to practical considerations, as the \bg \ wind emission is significantly reduced below this threshold at typical wind densities, and in particular drops off rapidly below 9000 K. The range at which the wind temperature could be reasonably treated as a free parameter of the model is thus narrow to the point that we only consider this temperature. \\ \indent
The velocity profile of the wind can, as previously for the magnetosphere, be separated into a poloidal component, which defines the velocity along the open field line, and a perpendicular component in the disk plane. The latter is largely driven by the fact that any field line, at its point of emergence from the disk surface, is effectively in Keplerian rotation about the stellar axis. Above the midplane, the rotational velocity of the gas stream deviates from a strict Keplerian profile and decreases with height and distance from the rotational axis.  In order to compute the local rotational component of the velocity, we follow \cite{Knigge1995} by assuming that the specific angular momentum with respect to the z-axis is conserved for any one stream line. 

The poloidal velocity in the direction of the gas stream is defined by a radial velocity law, controlled by the exponential parameter $\beta_{wind}$. This quantity effectively controls how quickly the wind reaches its terminal velocity as a function of radial distance from the centre, as well as of horizontal distance of the stream line anchor point at the midplane. The terminal velocity itself is defined as a factor f of the local escape velocity at the point of emergence. In our models, we again follow \cite{Knigge1995} and keep f=2 constant. \\ \indent
The gas density profile of the wind is computed from the local mass loss rate per unit surface area and the geometrical configuration of the wind. The local mass loss rate $\dot{m}$ itself is proportional to the temperature profile of the disk:
\begin{align}
    \dot{m}(R) \propto T(R)^{4\alpha}.
\end{align}
The exact nature of the temperature profile is not relevant in this context, as the continuum disk is not part of the model and the gaseous part is always below the cutoff height $z_{crit}$. The $\alpha$ parameter on its own, however, allows us to adjust the local mass loss, by which we can influence the radial distribution of the wind emission to some degree.
The exact proportionality between horizontal distance and local mass loss is determined from normalising the profile so that integrating the local mass loss rates of all the wind stream lines between the inner and outer radius adds up to the global mass loss rate $\dot{M}_{loss}$.  The density distribution is then fully described by the wind geometry, the velocity field, the global mass loss rate and the alpha parameter.
%
\\ \indent
In addition to the cutoff height $z_{crit}$, we lastly introduce a second new parameter that is not part of the original model description in \cite{Knigge1995}. The dark disk radius R$_{dd}$ defines the distance from the central star at which the midplane of the cell grid becomes completely opaque to all photons, effectively creating a thin black layer that blocks all radiation from the other side of the disk. This 'dark disk' extends from R$_{dd}$ to the edge of the cell grid, while any cell within the radius is completely transparent to all radiation. 
This again is very much a simplified approach to approximate the real influence of optically thick dust at certain distances in the midplane, but is sufficient to allow us to modulate the proportion of radiation from the back side of the disk that is going to contribute to our profiles. \\ \indent We finally note that the disk wind model is axisymmetric and thus effectively two-dimensional. While it is technically possible to combine a 2D wind with a 3D non-axsisymmetric magnetospheric
accretion model, we point out that such an approach would ignore the interactions between wind and accretion that arise in numerical treatments of such combined systems. In such a scenario, we would expect an azimuth dependence in the density of the wind and thus also an azimuth dependence of z$_{crit}$, which we do not consider in our analytical approach. 
In summary, we define the disk wind model component by setting the stellar parameters $R_*$, $M_*$, $P_*$
, the \textbf{global mass loss rate} $\dot{M}_{loss}$, the exponent regulating the \textbf{local mass loss per unit area} $\alpha$, the exponent regulating the \textbf{radial velocity law} $\beta_{wind}$,  the \textbf{inner radius} $R_{wi}$ and \textbf{width of the wind} $\delta R_{wind}$, the \textbf{dark disk radius} $R_{dd}$, the \textbf{focal point displacement} d, and the \textbf{temperature cutoff} z$_{crit}$. 

\subsection{Synthetic interferometric observables}
\label{sec:interfero}
The channel maps computed with MCFOST were used to extract synthetic spectral and interferometric data. The model spectral data can be computed straightforwardly by integrating the individual pixels of the brightness distribution over the entire image at each wavelength. Doing so yields a spectrum at the previously defined spectral sampling of the model, which contains the \bg \ line emission of the model components and a level of continuum emission defined by the stellar component. If the parameters set the gas in either accretion or wind component to be particularly hot and dense, this may add an additional smaller gaseous continuum contribution. \\ \indent
As the goal is to compare the synthetic to the observational data set obtained with GRAVITY for RU Lup in 2021, the synthetic spectrum needs to be degraded to the correct spectral resolution. This is achieved via convolution with a Gaussian kernel and interpolation on the observational wavelength grid. After normalising the spectrum to the continuum level of the image, this total line-to-total continuum flux ratio $F_{L/C}^{Im}$ in each channel is then modified to account for contributions coming from the dusty disk and the overresolved halo component to the total continuum which are present in the GRAVITY data but which we do not include in the model image:
\begin{align} \label{eq:obslineratio}
    F^{Obs}_{L/C}=\frac{F_{L/C}^{Im}+F_{IREx}}{1+F_{IREx}},
\end{align}

where $F_{IREx}=\frac{F_{Disk}+F_{Halo}}{F_{*}}$ is the infrared excess caused by the interferometrically partially resolved component $F_{Disk}$ and the overresolved halo component $F_{Halo}$. For the RU Lup 2021 data, we set the fraction of disk flux to the total continuum to 30 \% and the fraction of halo flux to 12 \% in accordance with the results reported in Section \ref{sec:Obs}. \\ \indent
The synthetic interferometric observables can be obtained from the Fourier transform of the image. A set of baselines mimicking the configuration of the GRAVITY UTs at the time of the observation were chosen to evaluate the Fourier integral at the corresponding points in the uv plane. In this manner we computed the visibilities and phases in each velocity channel, which were then also spectrally degraded akin to the treatment of the spectrum. We derived the sizes of the \bg \ emission region per channel from the synthetic visibilities in the same manner as from the observational data by removing the influence of the continuum first:
\begin{align} \label{eq:purelinevis}
    V_{Line}=\frac{F_{L/C}^{Im}V_{tot}^{Im}-V_{Cont}^{Im}}{F_{L/C}^{Im}}.
\end{align}

Here $V_{Line}$ is the pure line visibility, i.e. the part of the visibility from which the continuum contribution has been removed and which can be directly attributed to the line emission region. As Eq. \ref{eq:purelinevis} indicates, $V_{Line}$ does not require knowledge of the observational line-to-continuum ratio (Eq. \ref{eq:obslineratio}) and the associated continuum contributions. The modification based on disk and halo contribution is only strictly necessary for the comparison of synthetic and observational spectral data. \\ \indent
It is possible to relate the pure line visibilities to a characteristic angular size by presuming a certain morphology of the emission region brightness distribution, as was described in Section \ref{sec:Obs}. In this case, we follow the treatment of the \bg \ GRAVITY data by employing a simple Gaussian disk geometrical model to derive the characteristic size as the half width at half maximum, or half flux radius, of the radial brightness profile. Position angle and inclination of these centrosymmetric brightness distributions can be derived from such a model fit by taking into account the difference in size at different baseline angles. However, while this approach remains valid for disk-like structures, \cite{Tessore23} and \cite{Wojtczak2023} show that for the magnetospheric region close to the star the result of such a fit is difficult to interpret and does not immediately correspond to the physical inclination angle of the larger inner disk region. To remain consistent with our previous work and to ensure the best possible comparability, we remain with the same approach as before and fix the inclination of the system at the value obtained from UT data continuum fits, i=20$^{\circ}$. We compute the observables for a 90$^{\circ}$ position angle, as the uncertainty on the PA measurement is relatively high due to the low inclination of the system and the effect on the derived sizes is minimal. The stellar rotational axis in the model is then aligned with the north axis of the image and the disk would be aligned with the east axis. \\ \indent
The synthetic differential phases were computed by subtracting the continuum phase from the total line phase of the image. As the continuum is flat across the line, the continuum phase was taken as the average phase in the first and last image channel.  The continuum contribution was then removed to obtain the pure line differential phases, which is related to the angular offset on the sky plane via \citep{LeBouquin2009}
\begin{align}
    p_i=- \frac{\Phi_i \lambda}{2\pi B_i},
\end{align}

where $p_i$ is the magnitude of the offset along the i-th baseline as determined from the projection of the photocentre shift vector onto the baseline vector, $\Phi_i$ is the differential phase measured with that baseline in the $\lambda$ channel, and $B_i$ is the length of the baseline. The so computed angular offsets are relative in nature, they describe the positional shift between the barycentre of the \bg \ emission region relative to the barycentre of the underlying continuum brightness distribution. If the continuum region is centrosymmetric with regards to its emission, then the photocentre coincides with the position of the star and the differential phase yields the photocentre offset from the stellar position. While this is generally the case in the models, the observational data can potentially feature an additional continuum offset, caused by, for example, a local density fluctuation in the dust of the inner rim.
The magnitude and direction of the total photocentre shift vector can be determined by deprojecting the individual shifts measured along the six baselines of GRAVITY. For a more detailed description of this 
process, its technical steps and the derivation of the pure line quantity expressions, we refer to \cite{Wojtczak2023}.

\begin{table*}[t]
\footnotesize
\caption{RU Lup model configurations compared to VLTI GRAVITY data}
\centering
\begin{tabular}{ccccccc}	&	Mag. accretion I	&	Mag. accretion II	&	Disk wind I	&	Disk wind II	&	Hybrid model I	& 	Hybrid model II 	\\ &	Axisymmetric &	Non-axisymmetric &	Single peaked &	Double peaked	&	Cool + compact	& 	Hot + extended \\ &	(MA I)&	(MA II)&	(DW I)&	(DW II)	&	(HM I)	& 	(HM II)\\[0.07cm] \hline \hline \\
R$_*$ [R$_{\odot}$]	&	2.5	&	2.5	&	2.5	&	2.5	&	2.4	&	2.5	\\[0.1cm]
\rowcolor{mygray}M$_*$ [M$_{\odot}$]	&	0.8	&	0.8	&	0.8	&	0.8	&	1.2	&	0.8	\\[0.1cm]
T$_*$ [K]	&	4050	&	4050	&	4050	&	4050	&	4050	&	4050	\\[0.1cm]
\rowcolor{mygray}P$_*$ [days]	&	7	&	7	&	7	&	7	&	9.05	&	7	\\[0.1cm]
R$_{co}$ [R$_*$]	&	5.71	&	5.71	&	5.71	&	5.71	&	8.1	&	5.71	\\[0.1cm]
\rowcolor{mygray}R$_{mi}$ [R$_*$]	&	7	&	7	&	-	&	-	&	6	&	7	\\[0.1cm]
$\delta R_{m}$ [R$_*$]	&	2	&	2	&	-	&	-	&	1	&	2	\\[0.1cm]
\rowcolor{mygray}$\dot{M}_{acc}$ [10$^{-8}$ M$_{\odot} yr^{-1}$]	&	23	&	23	&	-	&	-	&	10	&	23	\\[0.1cm]
T$_{mag}$ [K]	&	8600	&	8600	&	-	&	-	&	7100	&	8600	\\[0.1cm]
\rowcolor{mygray}R$_{wi}$  [R$_*$]	&	-	&	-	&	7.1	&	9.1	&	7.1	&	9.1	\\[0.1cm]
$\delta R_{w}$ [R$_*$]	&	-	&	-	&	5	&	3	&	8	&	3	\\[0.1cm]
\rowcolor{mygray}d$_{fp}$ [R$_*$]	&	-	&	-	&	15	&	15	&	15	&	15	\\[0.1cm]
R$_{dd}$ [R$_*$]	&	-	&	-	&	14.5	&	13	&	13	&	13	\\[0.1cm]
\rowcolor{mygray}$\beta_{wind}$	&	-	&	-	&	0.62	&	0.05	&	0.2	&	0.05	\\[0.1cm]
$\alpha_{wind}$	&	-	&	-	&	0.4	&	0.7	&	0	&	0.7	\\[0.1cm]
\rowcolor{mygray}z$_{crit}$ [au]	&	-	&	-	&	0.0516	&	0.02	&	0.005	&	0.02	\\[0.1cm]
M$_{loss}$  [10$^{-8}$ M$_{\odot} yr^{-1}$]	&	-	&	-	&	5.1	&	12	&	10	&	12	\\[0.1cm]
\rowcolor{mygray}$\beta_{ma}$ [deg]	&	0	&	30	&	-	&	-	&	0	&	0	\\[0.1cm]  
\hline \hline \label{tab:ModelOverview}
\end{tabular}   
\tablefoot{Descriptions of the physical nature of the respective model parameters are given in Section \ref{sec:Macc} for the accretion model and in Section \ref{sec:dwmodel} for the disk wind model. Note that R$_{co}$ is implicitly defined by the stellar parameters.}
\end{table*}

\section{Comparison to GRAVITY RU Lup data}

\label{sec:results}

In this section, we present a number of different models in an attempt to recreate the observational profiles obtained from the GRAVITY data for RU Lup from 2021. As relevant observables we consider here the continuum-normalised line profile, the characteristic emission region size obtained by fitting a geometric Gaussian disk model to the visibilities, and the photocentre shift of the emission region as reconstructed from the differential phases. \\ \indent
Exploring the possible model variations is challenging given the large potential parameter space and the computational demands of MCFOST. Even a basic large scale model grid search over the entire parameter range is currently not possible within the technical capacities available and would require a substantial additional optimisation effort that is beyond the scope of this work. Instead, we utilised a mixed approach. We first considered the effect of isolated individual parameter variations from a common reference model  to identify those quantities with the strongest response. This analysis indicates that there is a clear distinction between parameters which influence primarily the line flux, but have minor impact on the interferometric quantities, and those that strongly affect also the characteristic size and photocentre shift. We refer to Appendix \ref{sec:sensitivity} for a more thorough discussion of those results. Second, we employed a semi-manual fitting routine which involves running partial model grids for which we varied only a subset of the total model parameters simultaneously within a manually defined bracket of possible values. The best fitting models from those partial grids were then determined by $\chi^2$ minimisation. This process was then iterated on by computing a new partial grid with different parameter variations to check if the $\chi^2$ could be further improved. \\ \indent
Given that the instrumental error on the data points computed by the GRAVITY pipeline lie on the order of less than 2\%, even the reduced $\chi^2$ is typically very large. We relied on these quantities exclusively as relative indicators in order to rank different parameter combinations and models against each other and did not derive any statement on the global statistical significance of the models from them.
With the simplified nature of our models and the limitations on the parameter exploration in mind, we did not necessarily expect to find a model that would be a full quantitative match of the observational data. Instead, the idea was to determine whether the main trends derived from the observation could be reproduced in principle on an acceptable level and to this end we prioritised achieving a good visual fit over a more quantitative approach. \\ \indent 
In this section, we describe in more detail two variants each of a pure magnetospheric accretion model, a pure disk wind model, and of a hybrid combination of both models. Table \ref{tab:ModelOverview} summarises the parameters of the different models.

\subsection{Stellar parameters}
The photospheric contribution to the continuum emission was modelled as a blackbody which radiates at an effective temperature $T_{*}$ over a surface area defined by the stellar radius $R_*$. In addition, the stellar mass $M_*$ and stellar rotational period $P_*$ affect the velocities in the magnetospheric funnel flows and wind outflows. While in \cite{Wojtczak2023} we assume those parameters to be relatively well defined for the test case of RU Lup, a more thorough review of the literature shows a significant spread of reported figures for this object. Consequently, we chose to adapt a broader range of possible parameter values, treating them effectively as a semi-free parameter in our model exploration.  The values adopted for the six models presented in this section are listed in Table \ref{tab:ModelOverview}.

\subsection{Scenario 1: Pure magnetospheric accretion model}

\begin{figure*}[t!] 
\begin{minipage}{0.48\linewidth}
    \centering
    \includegraphics[width=0.79\linewidth]{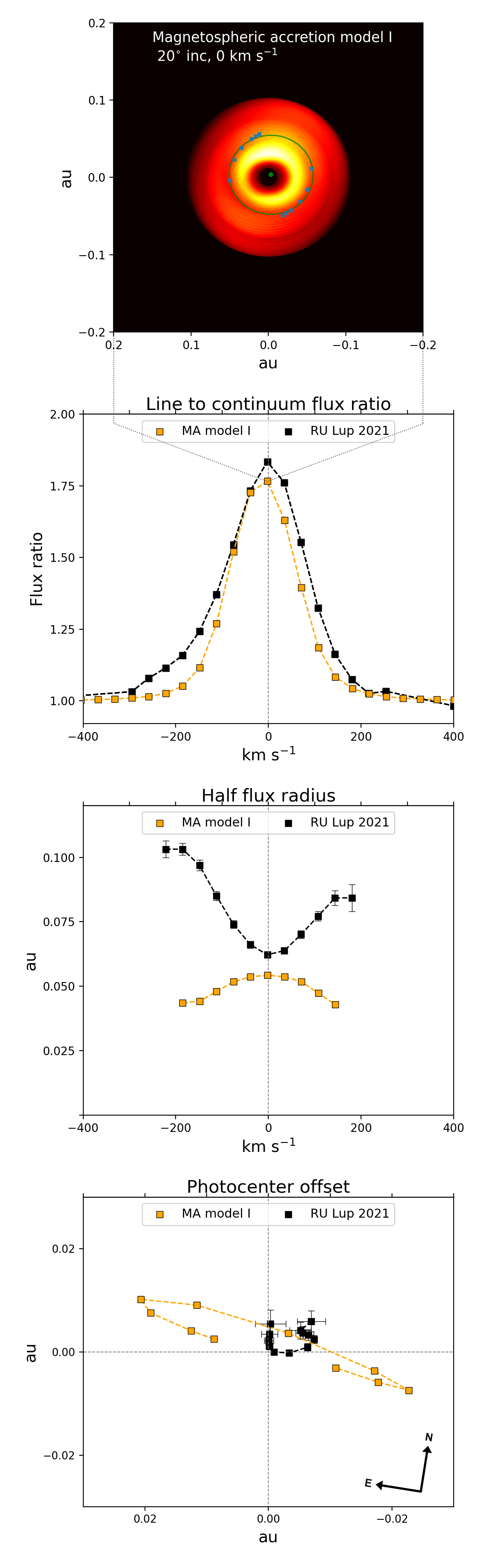}
\end{minipage}
\hfill
\begin{minipage}{0.48\linewidth}
    \centering
    \includegraphics[width=0.79\linewidth]{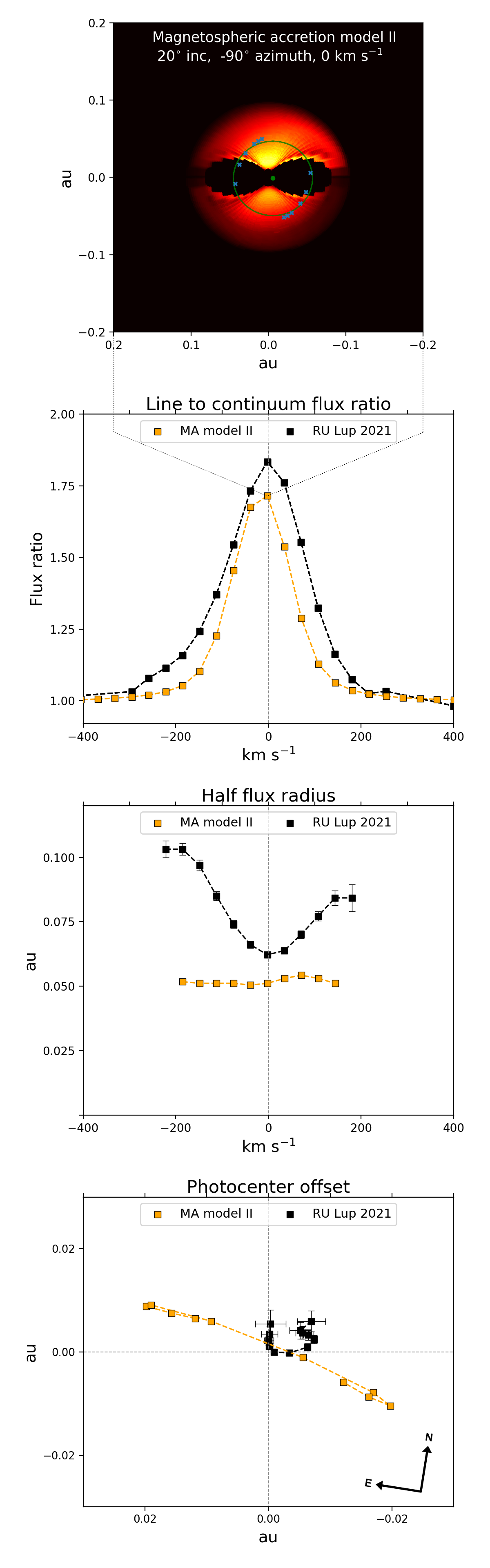}
\end{minipage}
    \caption{ Synthetic observables of the axisymmetric magnetospheric accretion model MA I \textbf{(left)} and the non-axisymmetric accretion model MA II \textbf{(right)} when viewed at an azimuth angle of -90$^{\circ}$. \\ \textbf{From top to bottom:} centre channel image, line-to-continuum flux ratio, characteristic size, and photocentre shift per velocity channel of the \bg \ emission region. The ellipse depicted in the image signifies the half flux radius of the geometric Gaussian disk model used to derive the characteristic size.  \\ In the depiction of the photocenter shifts, the stellar rotational axis, not the north axis, is aligned with the vertical coordinate axis. The observational photocentres were rotated by -9$^{\circ}$ with respect to Fig. \ref{fig:dataplot} to put them in the same frame of reference.
    } \label{fig:MA30OB20INC-90AZ}
    \end{figure*}
\begin{figure*}[t!] 

\begin{minipage}{0.48\linewidth}
    \centering
    \includegraphics[width=0.79\linewidth]{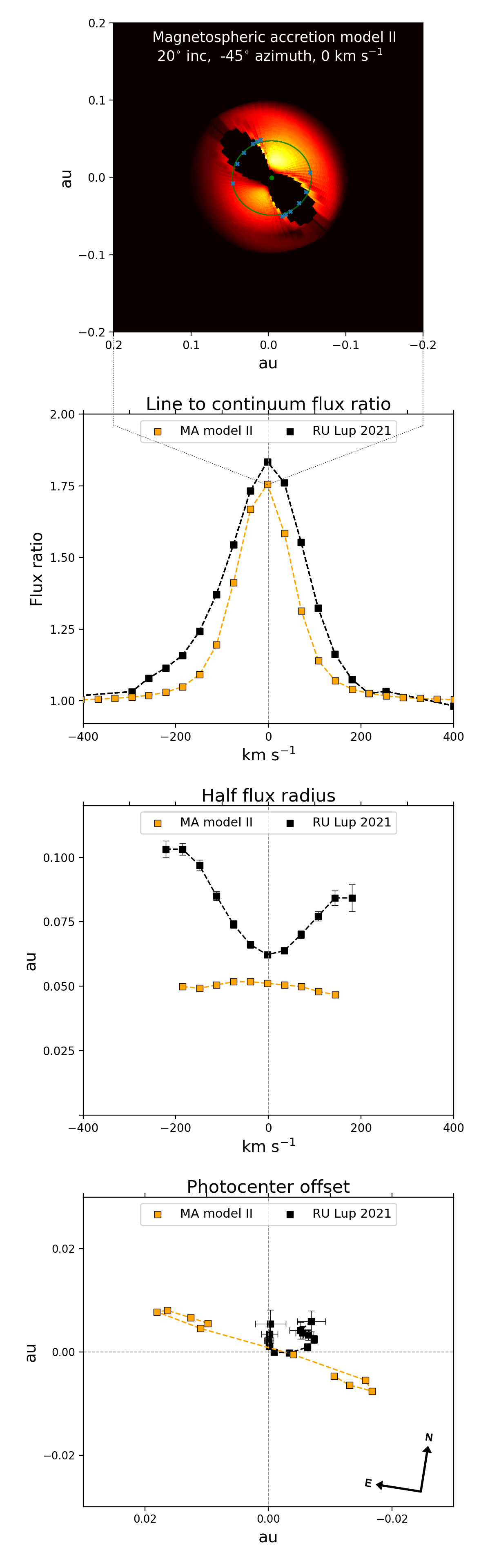}
\end{minipage}
\hfill
\begin{minipage}{0.48\linewidth}
    \centering
    \includegraphics[width=0.79\linewidth]{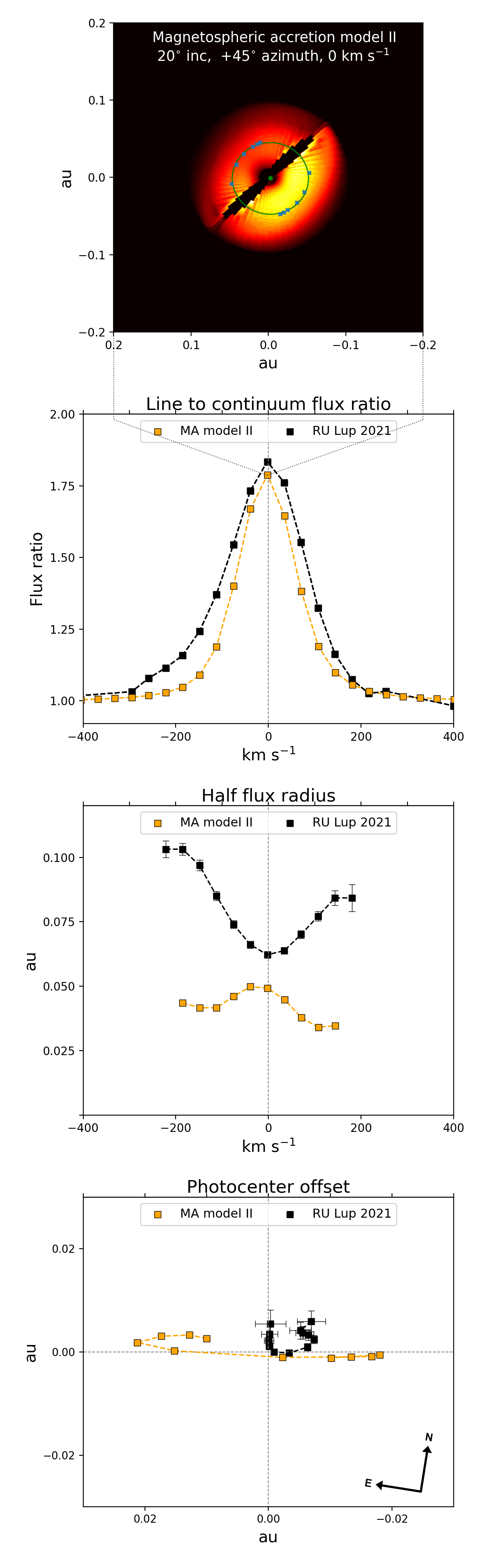}
\end{minipage}
    \caption{ Synthetic observables of the non-axisymmetric accretion model MA II when viewed at azimuth angles of -45$^{\circ}$ \textbf{(left)} and +45$^{\circ}$ \textbf{(right)}. \\ \textbf{From top to bottom:} centre channel image, line-to-continuum flux ratio, characteristic size, and photocentre shift per velocity channel of the \bg \ emission region. The ellipse depicted in the image signifies the half flux radius of the geometric Gaussian disk model used to derive the characteristic size.  \\ In the depiction of the photocenter shifts, the stellar rotational axis, not the north axis, is aligned with the vertical coordinate axis. The observational photocentres were rotated by -9$^{\circ}$ with respect to Fig. \ref{fig:dataplot} to put them in the same frame of reference.
    }
    \label{fig:MA30OB20INC-45AZ}
\end{figure*}

As a starting point, we revisit the case of the simple axisymmetric magnetospheric accretion model from \cite{Wojtczak2023}. In this work we explore a significantly expanded parameter space for this scenario, resulting in the model configuration depicted in the left column of Fig. \ref{fig:MA30OB20INC-90AZ} and referred to as MA I in Table \ref{tab:ModelOverview}. 
The normalised line flux is single peaked, with a centre channel line-to-continuum flux ratio of 1.76, compared to the observational peak flux ratio of 1.83. The line width between model and observation is comparable at the 50\% flux mark, with a model full width at half maximum (FWHM) of 168 km/s versus the observational profile at 187 km/s. At the base of the line, where the observational asymmetry is most pronounced, we note a larger disparity, with a width at 10\% flux (W10\%) of 320 km/s in the model versus a W10\% of 420 km/s in the GRAVITY data. Visually this difference is clearly localised in the blue wing of the line, where the model profile lacks the excess amount of blueshifted \bg \ emission that we detect in the observation. By contrast, the red wing of the line is well reproduced by the model.\\ \indent
In the centre channel, the fit of the Gaussian disk geometric model yields a characteristic size of 0.055 au. As such, the half flux radius sits at half the outer magnetospheric radius of 9 R$_*$, which at R$_*$ = 2.5 R$_{\odot}$ translates to 0.11 au. 
This result is again comparable to the observational centre channel half flux radius of 0.061 au, although still beyond the estimated 1$\sigma$ fit uncertainty of 0.001 au. The increase in characteristic size in the wings, however, which we see in the observational data, is not recovered by this accretion model. From the observation we derive a maximum size of 0.1 au at the extreme blue end of the wavelength range, while the model profile is centrally peaked and drops off to 0.045 au in the corresponding channel. \\ \indent
In the photocentre profile, the offset magnitude ranges up to $\pm$ 0.022 au, which is by about a factor of two larger than the largest photocentre shift we obtain observationally. The distribution of photocentres across the line has a similar 'crescent'-like shape, where the offsets at the most extreme velocities become smaller in magnitude again, but compared to the GRAVITY data the high velocity channels are still mostly aligned along a similar axis as the low velocity channels. The alignment of the overall model photocentre distribution appears almost perpendicular to the observational one, but the large uncertainty of the PA estimate obtained with GRAVITY ($\pm$ 30$^{\circ}$) leaves some ambiguity about the relative orientation. \\ \indent
Our second magnetospheric accretion model, referred to as MA II in Table \ref{tab:ModelOverview} and depicted in the right column of Fig. \ref{fig:MA30OB20INC-90AZ} for an azimuth angle of -90$^{\circ}$,
 addresses the question concerning the effect of a tilted magnetic dipole on the synthetic observables. Most of the fundamental parameters are unchanged with respect to MA I, but the model variant now features an obliquity of 30$^{\circ}$.
The change results in an overall drop in line flux
, with the peak flux reduced to 1.71 and the FWHM to 152 km/s. At an azimuth angle of -90 degrees
, this predominantly affects the red wing of the line profile. 
For the same angle, the centrally peaked shape of the size profile is flattened to a relatively constant size of 0.051 au across the line. A similar effect is seen for the photocentre profile, where the crescent arch is also compressed, although the maximum shift magnitude remains close to 0.022 au.  \\ \indent
The introduction of the dipole tilt breaks the axisymmetric nature of MA I. The image and resulting synthetic observables now depend on a selected azimuth angle, which effectively traces the rotation of the magnetosphere. At -90$^{\circ}$ in azimuth, the observer's line of sight is aligned with the two hemispheric funnel flows at non-zero inclinations. In Fig. \ref{fig:MA30OB20INC-45AZ} we concentrate on two specific examples at $\pm$ 45$^{\circ}$, which in conjunction with the -90$^{\circ}$ plot from Fig. \ref{fig:MA30OB20INC-90AZ} cover the relevant potential differences, while in Appendix \ref{sec:AzimuthAppendix} we present the full set of images for five different azimuth angles between $\pm$ 90$^{\circ}$. From these Figures it is clear that even qualitatively the nature of the impact of a dipole tilt depends on the azimuth. The localisation of the variation in flux ratio between axisymmetric and non-axisymmetric model changes between the red and blue wing as we move from negative to positive azimuth angles. The flattening of the size profile only appears at negative azimuths, while the positive orientations retain the centrally peaked shape. At +45$^{\circ}$ azimuth, the photocentre distribution is more strongly aligned with the disk axis, which signifies a shift of about 30$^{\circ}$ towards east relative to -90$^{\circ}$ azimuth and also relative to the axisymmetric case. \\ \indent
While the dipole tilt can clearly affect the overall shape of the observable profiles, its impact on the sizes and photocentre shift magnitudes at this level of inclination is even in the best case on the order of 10\% or less. In itself, this is not sufficient to fully resolve the limitations of the pure magnetospheric model in matching the GRAVITY data, as the discrepancies between model and data described for the non-axsisymmetric scenario largely remain the same.

\subsection{Scenario 2: Pure disk wind model}
\begin{figure*}[t!] 

\begin{minipage}{0.48\linewidth}
    \centering
    \includegraphics[width=0.79\linewidth]{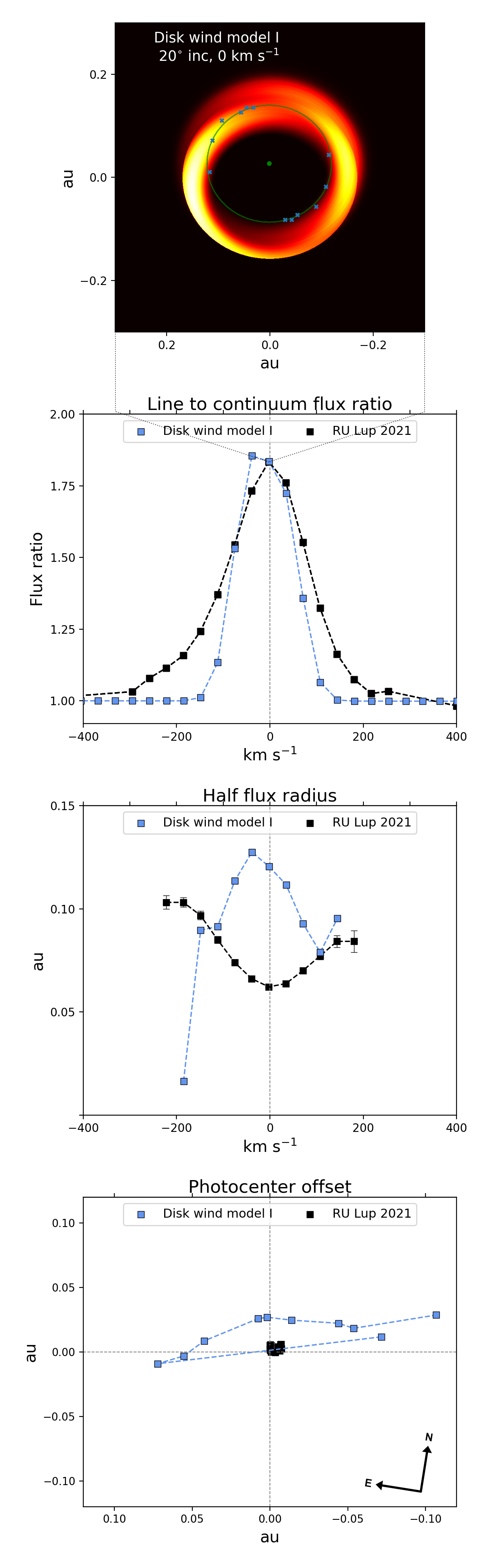}
\end{minipage}
\hfill
\begin{minipage}{0.48\linewidth}
    \centering
    \includegraphics[width=0.79\linewidth]{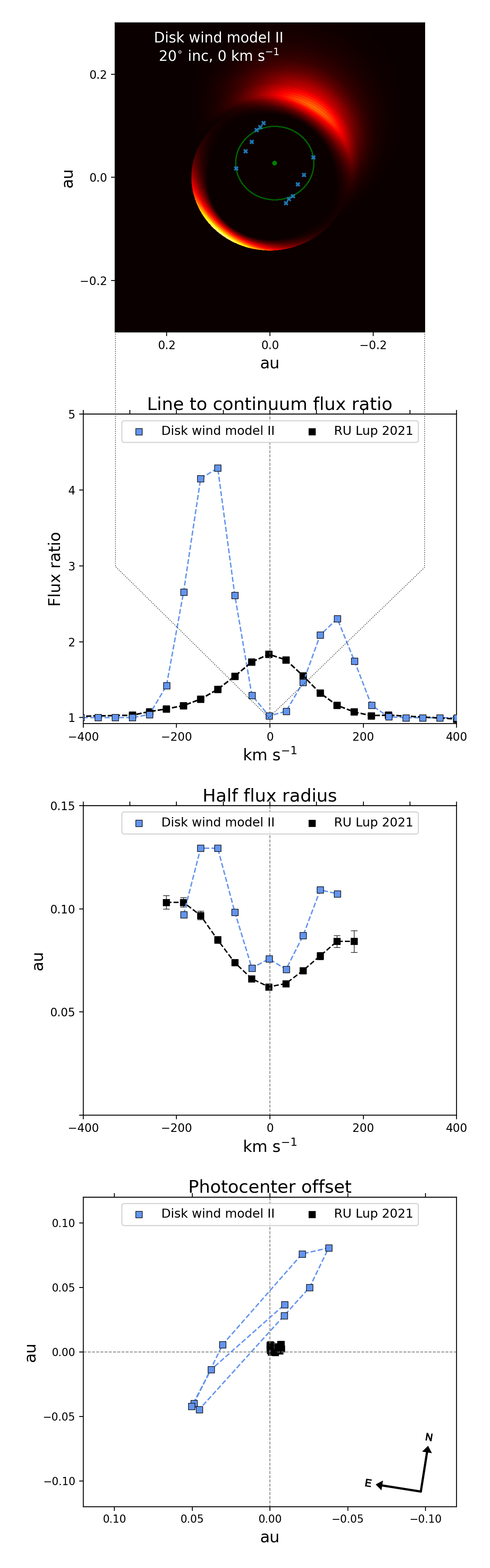}
\end{minipage}
    \caption{Synthetic observables of the slow disk wind model DW I \textbf{(left)} and the fast disk wind model DW II \textbf{(right)}. \\ \textbf{From top to bottom:} centre channel image, line-to-continuum flux ratio, characteristic size, and photocentre shift per velocity channel of the \bg \ emission region. The ellipse depicted in the image signifies the half flux radius of the geometric Gaussian disk model used to derive the characteristic size.  \\ In the depiction of the photocenter shifts, the stellar rotational axis, not the north axis, is aligned with the vertical coordinate axis. The observational photocentres were rotated by -9$^{\circ}$ with respect to Fig. \ref{fig:dataplot} to put them in the same frame of reference.}\label{fig:PureWind}
\end{figure*}

We further present two distinct pure disk wind models in order to explore the question  whether the RU Lup GRAVITY data might be better described by a disk wind rather than the magnetospheric accretion scenario. Both models, referred to as DW I and DW II in Table \ref{tab:ModelOverview} and depicted in Figure \ref{fig:PureWind}, are mainly differentiated by the wind velocity, as represented by a difference in the $\beta_{wind}$ parameter of 0.62 (DW I) to 0.05 (DW II). In practical terms, the differing wind velocities result in the line profile of DW I being single peaked, whereas the \bg \ line in DW II appears strongly double peaked. \\ \indent
The DW I model manages to replicate the central channel line flux ratio almost exactly at 1.83, although the line is more narrow with an FWHM of only 152 km/s and a W10\% of 231 km/s. The width at the base of the line in particular is then underestimated by a factor of almost two when compared to the W10\% of 420 km/s of the observational data.  \\ \indent
The size profile follows the centrally peaked shape of the \bg \ line itself, although the central half flux radius of the model overestimates the observational data by again a factor of two (0.121 au compared to 0.061 au, respectively). The large drop in size in the bluest channel in this case is a result of the low line flux at high velocities, which can threaten the integrity of the observable computation as the extraction of the pure line quantities requires division by a $(F_{L/C}-1)$ term. \\ \indent
The jumps in the photocentre positions at the most extreme velocities are likely equally caused by this effect, although we note that even at the centre channels, the shift magnitude is on the order of 0.025 au and thus almost three times as large as the most extreme observational offsets. At high velocities this discrepancy can grow to a factor of five, leaving the DW I profile much more extended than the observational profile. \\ \indent
\onecolumn


\begin{figure*}[t!] 

\begin{minipage}{0.48\linewidth}
    \centering
    \includegraphics[width=0.79\linewidth]{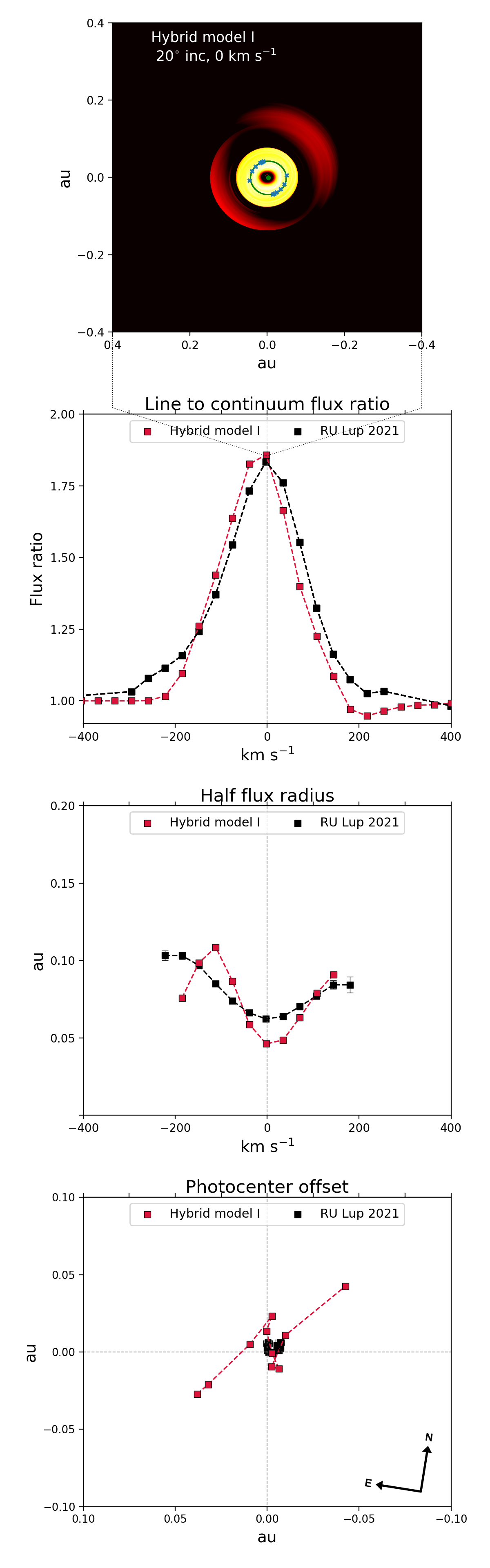}
\end{minipage}
\hfill
\begin{minipage}{0.48\linewidth}
    \centering
    \includegraphics[width=0.79\linewidth]{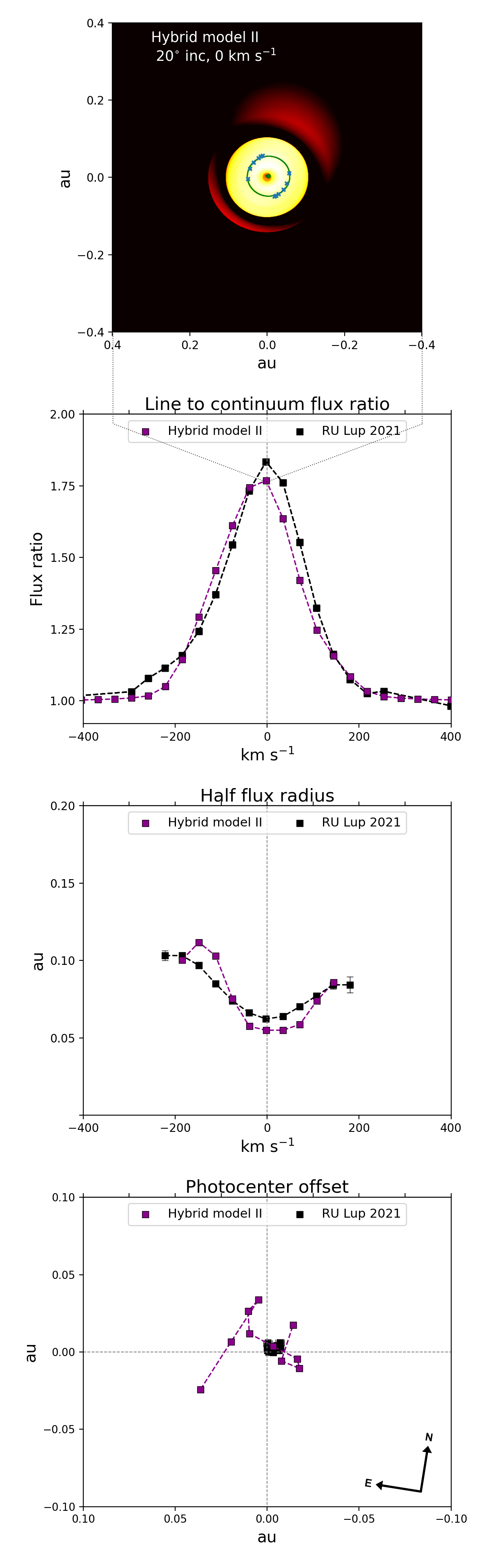}
\end{minipage}
    \caption{Synthetic observables of the cool, compact hybrid model HM I \textbf{(left)} and the more extended, hotter hybrid model HM II \textbf{(right)}. \\ \textbf{From top to bottom:} centre channel image, line-to-continuum flux ratio, characteristic size, and photocentre shift per velocity channel of the \bg \ emission region. The ellipse depicted in the image signifies the half flux radius of the geometric Gaussian disk model used to derive the characteristic size.  \\ In the depiction of the photocenter shifts, the stellar rotational axis, not the north axis, is aligned with the vertical coordinate axis. The observational photocentres were rotated by -9$^{\circ}$ with respect to Fig. \ref{fig:dataplot} to put them in the same frame of reference.}\label{fig:Hybrid1}
\end{figure*}
 \begin{figure*}[t!] 

\begin{minipage}{0.48\linewidth}
    \centering
    \includegraphics[width=0.79\linewidth]{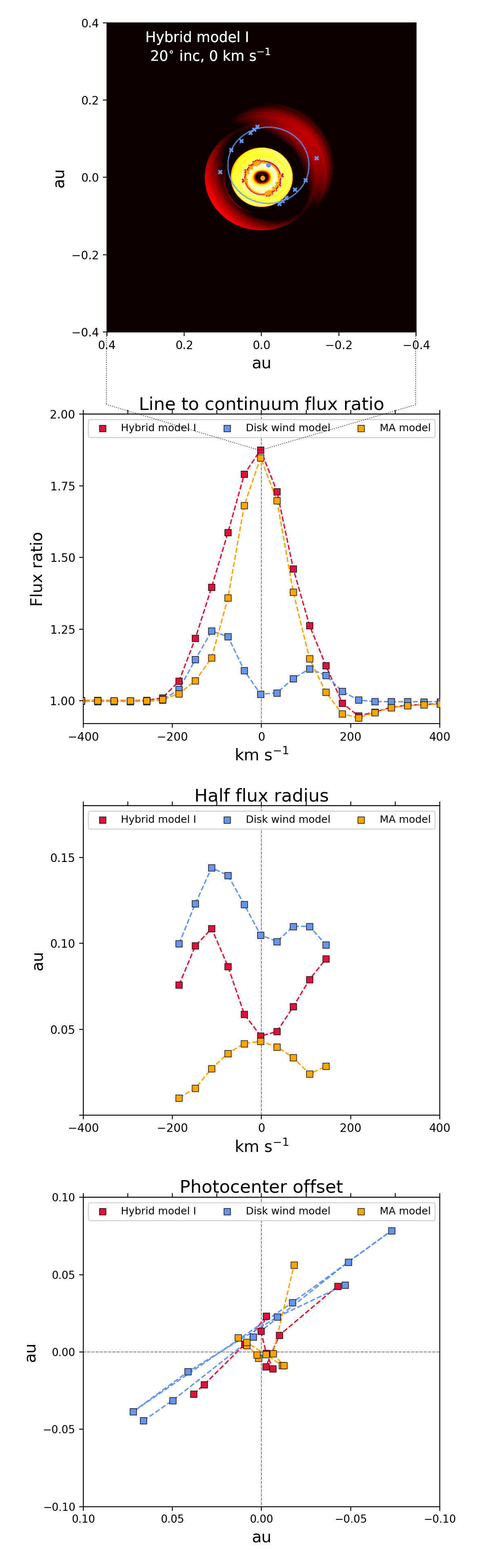}
\end{minipage}
\hfill
\begin{minipage}{0.48\linewidth}
    \centering
    \includegraphics[width=0.79\linewidth]{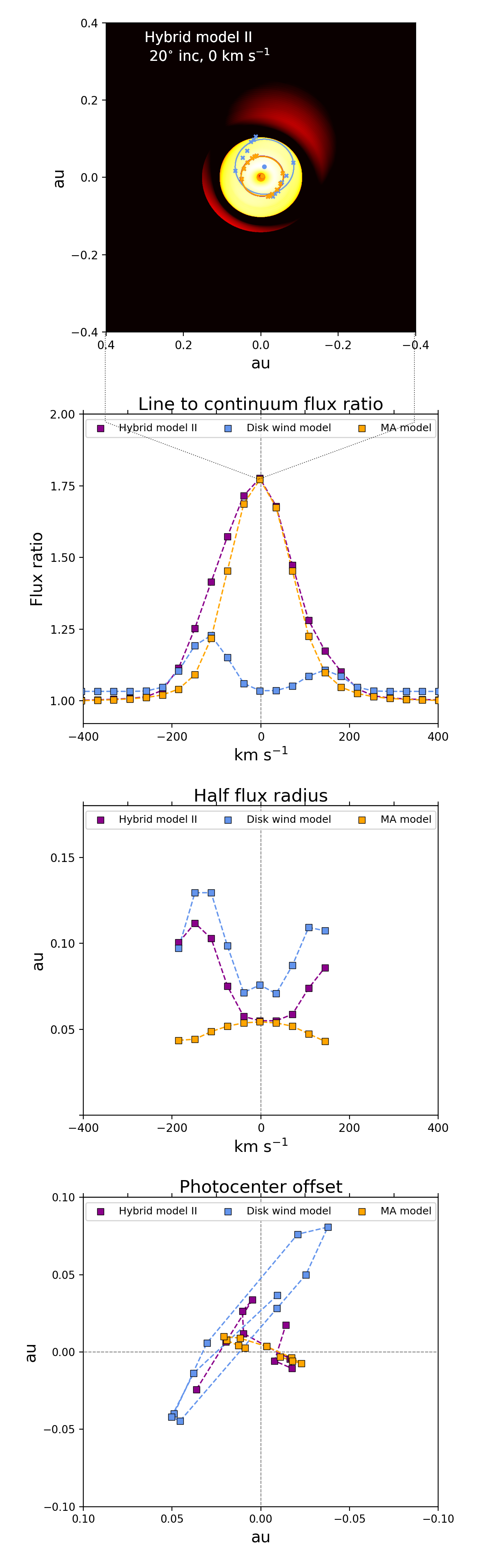}
\end{minipage}
    \caption{ Decomposition of the synthetic observables into their constituent model component profiles for the first \textbf{(left, HM I)} and second \textbf{(right, HM II)} hybrid model. \\ 
    In this depiction the stellar rotational axis, not the north axis, is aligned with the vertical coordinate axis. The observational photocentres were rotated by -9$^{\circ}$ with respect to Fig. \ref{fig:dataplot} to put them in the same frame of reference.} \label{fig:Hybrid2}
\end{figure*}
\twocolumn

The main purpose of the DW II model is to illustrate that the disk wind model can in principle reproduce the rising sizes at high velocities. The parameters of DW II were specifically selected to emulate the characteristic size profile of RU Lup as much as possible, which in this case comes at the expense of the line profile fit. The increase in size from a centre channel half flux radius of 0.07 au to a maximum size of 0.13 in the blue wing requires the line profile to be double peaked, which does not agree with the observational data. Additionally, while the centre channel sizes are close to the observational profile, those obtained in the wings of the profile are then larger by about 20\%.  The entire photocentre distribution is rotated by about 45$^{\circ}$ with respect to DW I, where the photocentres were mostly aligned parallel to the disk axis. Given that the shift magnitudes are comparable between both disk wind models, we find a similarly large disparity between model and observation for DW II as for DW I. 

\subsection{Scenario 3: Hybrid models}

Finally, we introduce a pair of hybrid scenarios in which we combine the disk wind and magnetospheric accretion region into a single common model brightness distribution. While both models, which in Table \ref{tab:ModelOverview} are referred to as HM I and HM II and which are depicted in Figures \ref{fig:Hybrid1} and \ref{fig:Hybrid2}, differ in a number of parameters, we first and foremost distinguish between them primarily based on the co-rotation criterion. For the model HM I, which is characterised by a cooler and more compact magnetospheric configuration than HM II, the size, rotational period and stellar parameters were selected to allow the magnetosphere to be truncated within the co-rotation radius. For HM II, we disregard this requirement and attempt to achieve the best fit to the data regardless of whether the co-rotation criterion is respected. \\ \indent
For HM I we find a centre channel flux ratio of 1.85 along with a FWHM of 180 km/s and a W10\% of 340 km/s. Flux ratio and FWHM deviate from the observational results by less than 5\%, although we see that W10\% is still underestimated. However, visually the disparity in the line base can no longer be clearly attributed to the lack of blue excess emission, but is also caused by the presence of an inverse P Cygni feature in the red wing which we do not detected observationally. On the contrary, the line flux in the blue wing is reproduced well up to velocities of about -150 km/s before it diverges significantly from the RU Lup data. \\ \indent
The centre channel characteristic size of 0.047 underestimates the observational result by less than 25\%, while the model still reproduces the increase in size towards the edges of the line. In the red wing the observational result is well matched at 0.09 au, although in the blue wing we do note a drop in size below -150 km/s, which does not agree with the data.  \\ \indent
The largest photocentre shifts at high velocities show a displacement magnitude of 0.045 au, which is still four times larger than the largest shifts observed for RU Lup. However, only the highest velocities show such a strong photocentre offset, whereas above -150 km/s the largest shift is already only about half as large (0.023 au). There is also a clearly visible change in the alignment of the profile at the same velocities. While the high velocity channels are aligned at an angle of about 45$^{\circ}$  relative to the disk axis, the remaining channels show photocentre offsets which are almost perfectly oriented along the stellar rotational axis. \\ \indent
The left column of Figure \ref{fig:Hybrid2} depicts the decomposition of the three model observables of HM I into their the constituent components. In the blue wing the wind becomes the dominant contributor to the line flux below -150 km/s, whereas in the red wing the influence is more mixed, as the inverse P Cygni feature of the magnetosphere still appears clearly in the combined line profile. The decomposition of the interferometric quantities shows that the shape in the wings largely follows the wind, but the combined region  characteristic size is adjusted downwards by the magnetospheric component. Equally, we see that the alignment of the photocentres at low velocities results from a superposition of differently aligned shift vectors between wind and magnetosphere. The high velocity channels follow the 45$^{\circ}$ alignment of the disk wind profile, although the magnetospheric influence again reduced the magnitude of the shift compared to the pure wind component. \\ \indent
For HM II, the centre channel flux ratio and FWHM (1.77 and 209 km/s, respectively) fit the observation similarly well as they did for HM I, but the W10\% is now increased to almost 400 km/s, which compares more favourably to the observational W10\% of 420 km/s. The line shape lacks the P Cygni characteristic of HM I, which improves the fit in the red wing to give an almost perfect match to the data points at velocities above +150 km/s. The flux ratio in the blue wing remains close to the observational ratio down to almost -200 km/s in this model. \\ \indent
The deviation in centre channel size is reduced to about 7\% and the sharp drop off in half flux radius at -150 km/s is much attenuated to provide an overall visually clearly improved fit to the data. 
In the photocentre shift profile of HM II, the largest offsets outside the extreme blue spectral channel are reduced in magnitude. In the red, the maximum offset is now 0.023 au, compared to the more than 0.06 au in the same channel of HM I. By contrast, the low velocity channels appear to feature photocentre shifts that are larger by up 20\% in magnitude with respect to their corresponding offsets in HM I and the axis of alignment does no longer coincide with the stellar rotational axis but is rather at a -45$^{\circ}$ degree angle relative to the disk axis. \\ \indent
From the decomposition of HM II shown in the right column of Fig. \ref{fig:Hybrid2} we see that the magnetospheric component in itself lacks the P Cygni characteristic and that it produces a line that is about 25\% broader when compared to HM I, while the relative contribution of the wind component is similar. The characteristic sizes of both wind and magnetospheric component are closer to the combined profile than before and the dominance of the magnetosphere in the centre channels is even greater as now up to three channel total characteristic sizes are almost fully defined by the magnetospheric component. This is again equally true for the low velocity channels of the photocentre shift profile, where the decomposition shows that the alignment of the total profile is also close to identical to the alignment of the magnetospheric photocentre offsets. 
A plot showing the mass density, temperature and velocity fields of HM II is included in Appendix \ref{fig:gridprofiles}.

\section{Discussion} 
\label{sec:discussion}
In Section \ref{sec:results} we present results for a total of six different model combinations, featuring two variants each of a pure magnetospheric accretion model, a pure parametric disk wind model, and a hybrid combination of both. The synthetic spectro-interferometric observables derived from these model images indicate that neither the magnetospheric accretion model nor the disk wind model are individually capable of reproducing the observational interferometric data obtained for RU Lup in 2021 with VLTI GRAVITY. While it is to some degree possible to approximate either the GRAVITY spectrum or the characteristic sizes of the emission region with specific components and parameter settings, a combination of wind and magnetosphere is ultimately required to reconcile model and observation. 
Both of the hybrid scenarios provide significantly better fits to the observational data and do not only reproduce the trends in principle, but come close to matching the normalised line flux and characteristic sizes. \\ \indent

\subsection{Non-axisymmetric magnetospheric accretion models}
The introduction of the dipole tilt has a significant effect on the geometry of the \bg \ brightness distribution, although the effective impact on the synthetic observables is tied strongly to its azimuth dependency. There is a qualitative change in the derived sizes across the line as we move from negative to positive azimuth angles (Fig. \ref{fig:MA30OB20INC-45AZ}, see also Appendix \ref{sec:AzimuthAppendix} for a larger range of azimuths). This behaviour is caused by the interplay between the system inclination and the position of the two hemispheric accretion flows relative to the observer's line of sight. At -90$^{\circ}$ azimuth, the upper hemisphere accretion column is inclined away from the observer, while at +90$^{\circ}$ azimuth it is inclined towards the observer. The latter case seems to favour the centrally peaked size profile, while the former shows that the sizes remain more constant at different velocities. However, the reason why the characteristic sizes are slightly larger at negative azimuth is not easily deduced, as there is also the baseline configuration to consider. While for the axisymmetric models the orientation of the baselines has a negligible impact, the azimuth dependency can change how their configuration probes the two hemispheric columns. \\ \indent
On the other hand, due to the nature of the photocentre shift computation, the effect of the baseline orientation on the derived offset position remains negligible even in the non-axisymmetric case. The differently oriented photocentre profiles at different azimuth angles can in this case then be directly attributed to the variation in the brightness distribution. There is once again a principal distinction between positive and negative azimuth angles, as the former lead to photocentre shift distributions that are more preferentially aligned with the disk axis, whereas the latter retain essentially the same angle of about 30$^{\circ}$ relative to the disk axis that was observed in the axisymmetric model.  \\ \indent  
While there is no obvious mechanism at play here that would directly explain why these models are associated with these specific alignments, the detection of different photocentre alignments at different azimuths is in itself noteworthy.
We observe a similar change in the orientation of the photocentre profiles between the 2018 and 2021 datasets. In addition, the observational size profile appears relatively more flat in 2018, whereas the 2021 data shows a stronger dip at low velocities (Fig. \ref{fig:dataplot}). The difference in centre channel size between both epochs is on the order of 0.02 au at comparable uv planes, which is significant compared relative to the uncertainty of 0.001 au on those points. It is also similar to the level of size variation we derive at different azimuth angles. We do note that this azimuth-dependent change in half flux radius primarily affects 
in the wings of the \bg \ feature rather than the centre. \\ \indent
Still, these azimuth-dependent effects may indicate that the two GRAVITY observations probe the RU Lup system at different points during its rotation. So while the switch to the non-axisymmetric dipole configuration does not resolve the fundamental limitations of the axisymmetric scenario with regards to fitting the RU Lup data, the inclusion of a tilted dipole offers one way to address the apparent time dependency of the observational results.

\subsection{Disk wind models}
The shortcomings of the pure wind models DW I and DW II are obvious, as they either fail to reproduce the behaviour of the characteristic sizes in the wings of the line or they produce a type of double peaked line profile inconsistent with the RU Lup data. The single peaked profile of DW 1 essentially replicates the fundamental characteristics of the magnetospheric accretion model. At the same time, it provides a worse fit to the data in terms of line width and especially photocentre shifts, thus offering no advantage over the magnetospheric accretion model. The double peaked configuration is more relevant, as it is the only type of low inclination model capable of matching trend of increasing sizes at high velocities. Indeed, DW II is identical in terms of model parameters to the disk wind component of the hybrid model HM II, where we successfully emulate the observational sizes in the wings by combining it with a hot magnetospheric central region. The line-to-continuum ratio depicted in Fig. \ref{fig:PureWind} appears much larger than in the decomposition in Fig. \ref{fig:Hybrid2} due to the lack of the magnetospheric continuum contribution, as will be discussed in Section \ref{sec:hybrids}. \\ \indent
 \begin{figure*}[t!] 
\begin{minipage}{1\linewidth}
    \centering
    \includegraphics[width=1\linewidth]{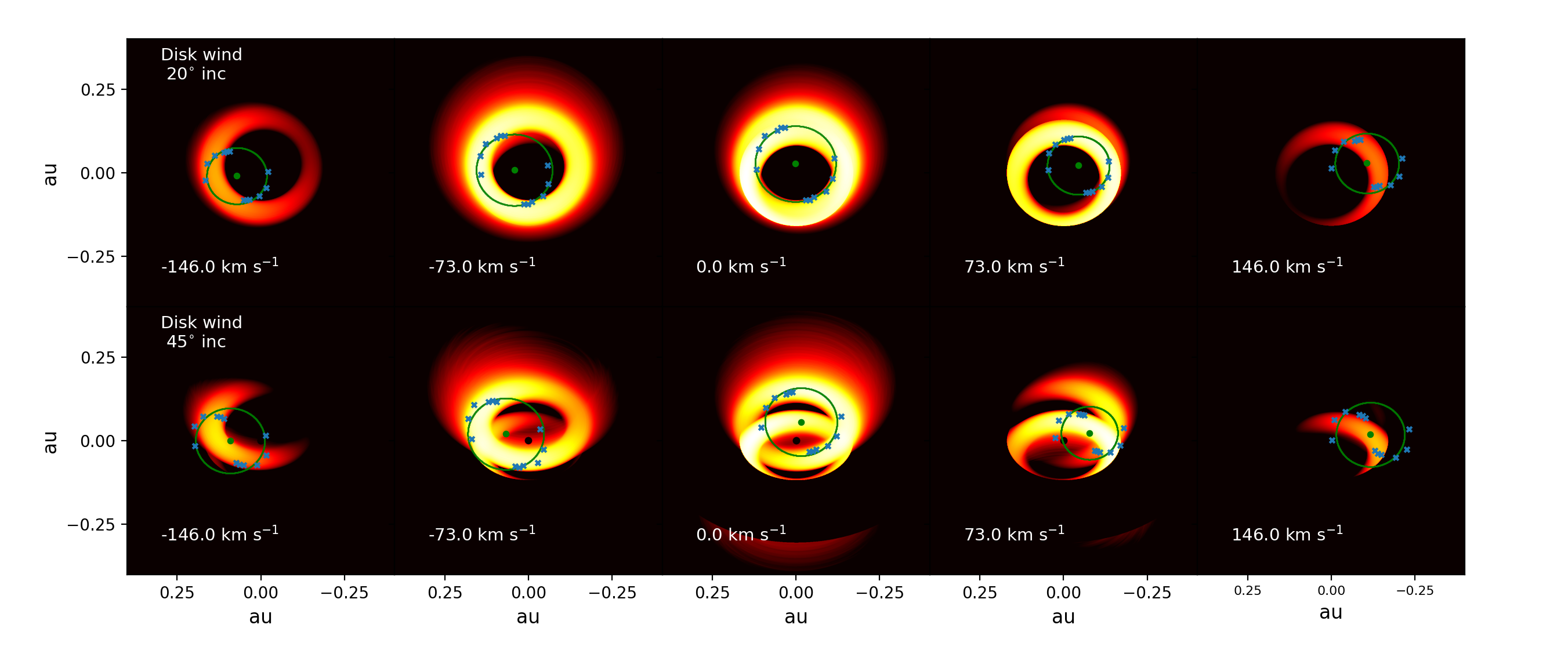}
\end{minipage}
    \caption{Images of the disk wind model DW I at different velocities across the \bg \ line and at 20$^{\circ}$ (top row) and 45$^{\circ}$(bottom row) inclination. The green ellipses depict the HWHM of the Gaussian disk model used to derive the characteristic sizes of the emission region. The Gaussian model is fixed at 20$^{\circ}$ inclination in both cases to remain consistent with the treatment of the original GRAVITY data. It is centred on the photocentre of each image, as recovered from the differential phases. The blue points signify the HWHMs obtained for the individual six baselines of the GRAVITY UT configuration that was used to observe RU Lup in 2021.}\label{fig:WindImages}
\end{figure*}
We remind that we consider the half flux radii obtained with the Gaussian disk model to be characteristic sizes first and foremost, and do not preoccupy ourselves with their exact relationship to the true physical size of the emission region for the purpose of this work. Still, it is worth pointing out for the sake of prudence that the disk wind model images essentially depict a ring-like region, which we fit with a disk-like geometric model. This may create a greater discrepancy between physical size of the region and characteristic size, especially in the wings of the line. Figure \ref{fig:WindImages} shows the HWHM of the Gaussian disk as an ellipse around the photocentre, compared to the underlying emission region. Especially at higher velocities, the relationship between both becomes more abstract, which may be a result of the mismatch in model morphologies. While this is of no concern to the comparison with the GRAVTIY data due to the consistent approaches between model and observational data treatment, a geometric ring model may better serve to derive values closer to the spatial extent of the true wind region.

\subsection{Physicality of the hybrid model parameters}
\label{sec:hybrids}
It is immediately obvious from Fig. \ref{fig:Hybrid1} that the hybrid approach is superior to any of the other models detailed in this work in terms of matching the observational data. It can both reproduce the general trends that we see in the observational line and characteristic size profiles and can, in the case of HM II, even fit the observational data reasonably well, given the limitations of the semi-manual fitting approach described in Section \ref{sec:Obs}. However, as the models are parametric in nature, there is a danger of enforcing a physically implausible parameter configuration in order to achieve a certain result. In this section we discuss the selected parameters in the context of the available literature on RU Lup. \\ \indent
While an examination of the physicality is difficult for parameters without explicit connection to quantifiable observables, there are those for which observational evidence or implicit constraints are more readily available. Chief among them are the stellar parameters, which are only partially shared between the hybrid models. For HM I and II the stellar mass was set to 1.2 M$_*$ and 0.8 M$_*$, respectively, the stellar radii to 2.4 R$_*$ and 2.5 R$_*$, and the stellar temperature to 4050 K. These masses are well within the broad range of possible values proposed by \cite{Alcala2017}. They estimate stellar mass and mass accretion rates from four different evolutionary models for pre-main sequence stars, thus yielding four different mass estimates between 0.43 M$_*$ and 1.21 M$_*$. They also report a stellar radius of 2.39 $\pm$ 0.55 R$_*$ and a photospheric temperature of 4060 $\pm$ 187 K.  \\ \indent
The mass accretion rate for HM I was set to 10 $\cdot$ 10$^{-8}$ $M_{\odot}\ yr^{-1}$ and for HM II to 23 $\cdot$ 10$^{-8}$ $M_{\odot} \ yr^{-1}$. This agrees with our own estimates of the accretion rate for RU Lup based on the GRAVITY \bg \ data. In \cite{Wojtczak2023} we use the empirical accretion luminosity to line luminosity relationship given by \cite{Alcala2014} to determine M$_{acc}$ = 13.38$^{+15.79}_{-6.66}$  $\cdot 10^{-8} \ M_{\odot} \ yr^{-1}$ for the 2021 data and 17.31$^{+20.78}_{-8.54}$  $\cdot 10^{-8} \ M_{\odot} \ yr^{-1}$ for the 2018 data of RU Lup.  The highest estimate available in the literature is found by \cite{Siwak2016} at 30 $\cdot 10^{-8} \ M_{\odot} \ yr^{-1}$. \\ \indent
Since the magnetospheric accretion model does not treat the gas temperature in the funnel flows self-consistently, one needs to be cautious to choose a maximum accretion flow temperature consistent with observations. Hard limits to the chosen combination of accretion rate and temperature can be easily enough determined by looking at extreme cases. At either very low mass accretion rates or temperatures ($\leq$ 6000 K, $\leq$ 10$^{-9} \ M_{\odot} \ yr^{-1}$) barely any \bg \ emission is produced and the image is almost completely continuum dominated. Increasing temperature and accretion rate simultaneously will lead to veiling effects as the gas starts emitting in the continuum and becomes increasingly optically thick. This behaviour was already observed in the magnetospheric accretion study of \cite{Muzerolle2001}. They found that, for all the investigated hydrogen transitions, a critical threshold was reached at temperatures $\geq$ 10000 K and accretion rates of $\geq$ 10$^{-6} \ M_{\odot} \ yr^{-1}$, at which point the emission line is turned into an absorption feature. At lower densities however, the gas never becomes sufficiently optically thick for this effect to occur. In their work they derive optimal ranges of mass accretion rates and temperatures for T Tauri stars from the continuum luminosity and the Pa$\beta$/\bg \ line ratios.  For our hybrid models, we paired the mass accretion rate of HM I (log(M$_{acc}$)=-7) with a maximum magnetospheric temperature of 7100 K and the accretion rate of HM II (log(M$_{acc}$)=-6.6) with a temperature of 8600 K. According to Fig. 16 in \cite{Muzerolle2001},  the excluded temperature ranges for log(M$_{acc}$=-7) are given as below 6500 K  and above 10000 K and for log(M$_{acc}$=-6.6) below  6000 K and above 9500 K. Our chosen accretion and temperature parameters for both hybrid models lie well within this range of possible, or not excluded, combinations. \\ \indent
\cite{Fang2018} show for a population of T Tauri YSOs, RU Lup among them, that $\dot{M}_{wind}$/$\dot{M}_{acc}$ ratios can be on the order of slightly above 1, depending on the gas temperature, with a median ratio of 0.1 derived from a number of atomic emission lines for their sample. This is similar to the findings of \cite{Watson2016} and is consistent with the global mass loss rate we set for the hybrid models, where we use $\dot{M}_{loss}$=$\dot{M}_{acc}$ for HM I and $\dot{M}_{loss} \approx 0.5 \ \dot{M}_{acc}$ for HM II. However, these values are derived for the low velocity component of the outflows,  with velocities on the order of a few tens of km/s. For the high velocity component, which is more in line with the velocities we require for our wind components, \cite{Fang2018} presents maximum ratios on the order of 0.1 or less. This would suggest that our wind component is significantly stronger than what has been observed in these studies. We also note that the high $\dot{M}_{wind}$/$\dot{M}_{acc}$ ratios found for the low velocity component are based on more extended wind regions than the ones used in our models. \\ \indent
It may be useful to discuss the dark disk radius, which in both hybrid models is set to 13 R$_*$, in the context of gas and dust in the innermost parts of the disk. Historically, the simplified view of inner disk opacities assumes that the close environment of a young star is divided into an inner optically thin purely gaseous disk and an outer optically thick dusty region, separated by the sublimation radius \citep{Natta2001}. Although some studies propose that the opacities of the gaseous cavity inwards of the sublimation radius behave in a more complex manner than this image suggests (see e.g. \cite{Muzerolle2004}), the gaseous inner disk is only likely to become optically thick if the accretion rate is sufficiently large. 
If we, in first order, assume that in our case the midplane within the inner cavity is close to transparent to \bg \ radiation, then the change in opacity would be predominantly connected to the presence of dust. In \cite{Perraut2021}, the K band continuum emission is traced to a region with a half flux radius of 0.21$^{+0.07}_{-0.05}$ au using a geometric Gaussian ring model. The general assumption is that this region is a proxy for the hot inner dust wall, which is thought to dominate the disk continuum emission at NIR K-band wavelengths. In the same publication, we compare the K-band size to a sublimation radius derived from a simple model of dust absorption and emission \citep{Monnier2002} and this way find a smaller value of 0.05 to 0.1 au.  At a dark disk radius of 13 R$_*$ (0.15 au), the disk in the hybrid models becomes transparent just in between these two regions.  \\ \indent
We set the rotation period in HM I to 9.05 days and in HM II to 7 days, which is consistent with known general distributions of rotational periods in low mass YSOs. These can range up to about 10 days, as was shown in a number of studies based on the data from the \textit{Kepler} space telescope (e.g. \cite{Rebull2020}). However, for the specific case of RU Lup, the rotational period has been estimated to be on the lower end of this dispersion.  \cite{Stempels2007} investigated the periodicity of radial velocity variations for this object and derived a stellar period of 3.71 days, consistent with earlier photometric periodicity studies, for example by \cite{Hoffmeister1965} and \cite{Drissen1989}. \cite{Frasca2017} use the VLT X-Shooter spectrograph to determine the v sin(i) of RU Lup to be 8.5 km/s, which translates to a stellar period of 5.1 days when using the inclination fitted from the GRAVITY NIR continuum data (20$^{+6}_{-8}$ degrees, see \cite{Wojtczak2023}). Even using the highest deviation of 26 degrees within the stated 1$\sigma$ range will only yield a rotational period of 6.5 days in this case. It is not clear why our best fit models point to stellar periods that are significantly slower than known observational results, but the question of the model stellar period is ultimately also tied to the size of the magnetosphere and thus to the accretion regime of RU Lup.

\subsection{The accretion regime of RU Lup}
The fundamental distinction between the two hybrid models HM I and HM II lies with how they treat the co-rotation criterion. HM I represents a parameter combination that fully respects the criterion: The sizes, stellar parameters, and rotational period were selected so that R$_{mo}$ $\approx$ R$_{co}$. The magnetosphere is then truncated on the order of the co-rotation radius R$_{co}$, which is consistent with the idea of stable magnetospheric accretion. 
In numerical MHD simulations, a further distinction arises between the regimes of stable (0.7 R$_{co}$ $\leq$ R$_{mo}$ $\leq$ R$_{co}$) and unstable (R$_{mo}$ < 0.7 R$_{co}$) accretion \citep{Romanova2015}. If the magnetospheric truncation is more compact, then the disk can penetrate into the magnetosphere and the gas is accreted from close to the star along a set of spontaneously forming accretion tongues.
Additionally, for R$_{mo}$ > R$_{co}$, angular momentum considerations dictate that the accreting matter would be at least partially ejected at the disk-magnetosphere interface and thus lead to less steady, more time-variable accretion \citep{Blinova2016}. This is referred to as the 'propeller regime'.  In \cite{Romanova2015}, a 'fastness parameter' $w_s=\left(\frac{R_{mi}}{R_{co}} \right)^{(3/2)}$ is used to distinguish between different accretion regimes and also between strong and weak propellers. They define a system in a weak propeller regime as one where $w_s \approx 1$, while a w$_s$ at multiples of 1 indicates a strong propeller with ejection exceeding the remaining accretion. \\ \indent
For HM II, we decided to relax the co-rotation criterion and set the truncation radius and the stellar period independently to obtain the best possible fit to the GRAVITY data. The chosen truncation radius of 7+2 R$_*$ compares to a co-rotation radius of 5.72 R$_*$ at the selected P$_{rot}$ = 7 days. Our configuration of HM II translates into a fastness parameter of 1.35, which would indicate that - if we straightforwardly apply these concepts to our analytical model -  it still exists in the regime of a weak propeller.  HM II could then still be consistent with a system that is principally in accretion, even if some material is ejected at the magnetospheric boundary. Such a scenario of non-steady accretion in a weak propeller regime was discussed in  \cite{Zanni2013}, see also \cite{PantolmosZanni2020}.
\\ \indent
However, this would contradict the idea that RU Lup is in a regime of unstable accretion, as proposed by \cite{Siwak2016}. 
Their examination of RU Lup photometric data from MOST observations taken in 2012 and 2013 show highly irregular light curves for this object. Based on this lack of a clear periodicity in the brightness variability, they argue that the chaotic behaviour is the result of a randomised distribution of temporary accretion hot spots on the stellar surface caused by the unstable accretion tongues. This is in agreement with the theoretical predictions of unstable accretion light curves from numerical simulations \citep{Romanova2008-2}, and indeed this regime of unstable accretion is associated predominantly with higher accretion rates such as found for RU Lup. It is not clear whether the remaining accretion in a system in the propeller regime might produce a similar effect.\\ \indent
Another question as to whether the analytical stable accretion model yields plausible physical characteristics of the magnetic field. While we do not set a magnetic field strength explicitly, the chosen stellar parameters, mass accretion rate, and truncation radius can be implicitly related to a magnetic dipole field strength. Using the formulation in \cite{Hartmann2016}, we can determine the field strength in kG via
\begin{align}
    B_{kG}=\left( \frac{R_{tr} M_{0.5}^{(1/7)} \dot{M}_{-8}^{(2/7)}}{18\gamma R_{2}^{(12/7)} } \right)^{(7/4)},
\end{align}
where M$_{0.5}$ is the stellar mass in units of 0.5 M$_{\odot}$, $\dot{M}_{-8}$ the mass accretion rate in units of 10$^{-8} \ M_{\odot} \ yr^{-1}$, R$_2$ the stellar radius in units of 2 R$_{\odot}$, and $\gamma$ is an uncertain factor $\leq$ 1. Assuming the most favourable case with $\gamma$ = 1, and using the inner magnetospheric radius as truncation radius (i.e. 6 R$_*$ for HM I and 7 R$_*$ for HM II), the field strength for HM I comes out to 1.5 kG, whereas for HM II we compute a 2.6 kG dipole field. Although there are no published measurements of the RU Lup field in the literature, a default assumption for the magnetic fields of T Tauris stars is a range of about 1 to 2 kG, so the implicit field strength of the HM II configuration would be atypically large. We do note, however, that this formula is approximate and based on simplifications of the mechanics involved. If indeed the accretion process in RU Lup is complex beyond the case of simple stable magnetospheric accretion, as the previous considerations imply, then these magnetic field values should not be adopted without question. \\ \indent
The question of the RU Lup accretion regime is ultimately not easily answered. The stable analytical accretion model is also only an approximation of a more complex truth, and the fact that some of the parameters we derive from it are not immediately consistent with other observations is not unexpected. Another example for a potential additional effect that is not considered at all outside certain numerical simulations is the idea of a failed disk wind, meaning a wind that is launched close the disk-magnetosphere interface, but which fails to escape the gravitational field of the star. Instead, the wind material falls back onto the magnetosphere and is accreted onto the star, effectively enlarging the magnetospheric emission region and masking a more compact actual magnetosphere in the process \citep{Takasao2022}. \\ \indent
A different  potential explanation for the size of the best fitting magnetosphere may lie with the treatment of the observational data rather than with the models themselves. Contrary to the treatment of the continuum data in \cite{Perraut2021}, the visibility modelling of the \bg \ line emission data did not consider the effects of a large scale halo component that might account for a significant part of the line flux. Such a halo component, which may for example reflect the influence of \bg \ photons scattered on the disk surface, was omitted from the treatment as the primary intent behind \cite{Wojtczak2023} was to compare general trends among a number of objects for a simple geometric approach. Including a halo would effectively reduce the derived characteristic size for the resolved emission region compared to the results presented in Fig. \ref{fig:dataplot}. Whether this effect could be sufficiently strong to reduce the observational characteristic sizes to a point where we may be able to fit them similarly well as we did with HM II, without exceeding the co-rotation radius, is at this point entirely speculative and would depend on a number of other factors. Still, it may be a point worth considering for future investigations into similar questions.


\subsection{The photocentre shift problem and the possible role of other emission components}
One of the primary issues of the model analysis in \cite{Wojtczak2023} is that the synthetic photocentre positions obtained from the images are offset by a factor four to five when compared to the GRAVITY photocentre shifts. Neither the consideration of a dipole tilt, nor the introduction of an additional wind component, do much to resolve this issue. On the contrary, the photocentre profile associated with the disk wind shows even substantially larger photocentre shifts than the magnetosphere. As a consequence,  the hybrid photocentre profiles become even more elongated, as shown via the decomposition in Fig. \ref{fig:Hybrid2}.  \\ \indent
The divergence between the observational and synthetic profiles is potentially rooted in some aspect of the observed system that is not reflected in the models, such as the presence of an additional \bg \ emission component. Such a component could, for example, come in the form of  a hot MHD stellar wind, launched along the open magnetic field lines from the polar regions of the star \citep{Ferreira2006}. However, it is not clear whether such a stellar wind would even address the issue of the photocentres in principle, let alone be consistent overall with the GRAVITY data. In a low inclination system such as RU Lup, a collimated outflow, launched from the stellar pole, almost directly along the line of sight would be expected to cause significant blueshifted absorption, whereas GRAVITY on the contrary picked up blueshifted excess emission in the \bg \ spectrum. Given that the redshifted stellar wind component is also likely to be at least partially blocked by the star itself at this inclination, there should be a substantial difference in both characteristic size and photocentre shift profiles between the red and the blue arm. It is equally not clear whether such a collimated wind could simultaneously reproduce the increase in size in the wings and the more compact distribution of photocentres. \\ \indent
Any component that may help to resolve the question of the overestimated photocentre shifts fundamentally needs to lead to a decrease in offset magnitudes without an accompanying reduction in characteristic size.  This would be achieved by a system that remains relatively centrosymmetric across the line at scales even up to the truncation radius, but then features asymmetries in the extreme close environment of the star. A disk-like structure that would extend to the innermost regions of the system, viewed at low inclination,  may provide the necessary degree of centrosymmetry across the line, but lacks the asymmetric elements. In this context we could consider again the possibility that RU Lup may be in the regime of unstable accretion. If the gaseous matter penetrating the magnetosphere behaved essentially as such a disk, then the formation of accretion tongues very close to the star could introduce the necessary asymmetry that would lead to offsets in the photocentre positions. Answering this question would require some dedicated modelling, as without knowledge of the relative \bg \ flux between the tongues and the remaining emission region, it is difficult to comment on whether this would be a viable solution. This is also true for the introduction of a halo component, which is another possible way to address this question. The scattered light halo may in principle have a similar compressing effect on the photocentre shifts, assuming that it itself is largely centrosymmetric. Whether this effect could be sufficiently large to explain the degree of divergence we detect between model and observation is again a question of the flux ratios between halo and resolved region.  Still, as of now, these ideas may hold the most promise to resolve the issue of the photocentre shifts.

\section{Summary}

We computed the spectro-interferometric characteristics of six models of the \bg \ line emission region in the innermost environment of low mass young stars.  These models featured different configurations of emission components in the form of disk wind outflows and magnetospheric accretion.  From the synthetic model data, which included the normalised line-to-continuum flux ratio, the interferometric visibilities, and the differential phases, we determined characteristic sizes and the position of the emission region photocentres for ten spectral channels across the \bg \ line. We then compared these synthetic quantities to observational results derived from VLTI GRAVITY data for the classical T Tauri star RU Lup.\\ \indent
 Our goal was to explain a number of features in the observational data that were not well reproduced by previous modelling efforts presented in \cite{Wojtczak2023}. Specifically, we tried to determine whether the addition of a larger scale outflow component onto an accreting magnetosphere would deliver results consistent with an asymmetric line profile with blueshifted excess emission, an increase in characteristic radii towards the high velocity wings of the line profile, and a compact distribution of photocentres across the line.
 To summarise the most important results of this paper:
\begin{itemize}
\item{Both axisymmetric and non-axisymmetric magnetospheric accretion models produce either centrally peaked or flat characteristic size profiles across the line. No parameter configuration leads to a size increase at high velocities and a minimum size at the line centre, as observed for RU Lup. While it is possible to fit the central component of the \bg \ line with a hot magnetosphere, the blueshifted excess emission that we detect observationally cannot be emulated.}
\item {The variability in equivalent width, peak flux ratio, and photocentre shifts that we find between the 2018 and 2021 epochs of the observation is on the order of the variation that we detect when comparing the line profile of the non-axisymmetric magnetospheric accretion model at different azimuth angles. This could indicate that between 2018 and 2021 we are in observing a tilted dipole at different phases during the stellar rotational period.}
\item {Only disk wind models with sufficiently fast winds to produce double peaked spectral features at the spectral resolution of GRAVITY are associated with an increase in emission region sizes at high velocities. We do not manage to reproduce both the centrally peaked line profile and the increase in the size profile simultaneously with a disk wind alone.}
\item {Hybrid models, combining a strong magnetospheric component that dominates the central channels of the line and a fast double peaked wind component, can reproduce the trends in both line profile and sizes at the same time as parameters that appear mostly in agreement with previous observational findings on the physical characteristics of RU Lup. The magnitude of the model photocentre shifts overestimates the observational photocentre profile across all model variations by factor of four or more, including the hybrid models.}
\item{Our best fitting hybrid model manages to well reproduce the 2021 \bg \ spectral line and characteristic sizes across the line.  Still, in that case, the truncation radius exceeds the co-rotation radius by more than 50\%, meaning that part of the gas is found beyond the centrifugal barrier, where it is expected to be expelled and not accreted.}
\end{itemize}

In conclusion, we find that the success of the hybrid approach to modelling the \bg \ region strongly supports the idea of a multi-component emission environment for certain class II low mass YSOs with large mass accretion rates such as RU Lup. Discrepancies between the expected magnetospheric accretion characteristics and our best fitting parameters likely indicate that the accretion-ejection process in RU Lup is more complex than what is captured by our analytical models. Another explanation may lie in the potential effects of a possible unresolved  \bg \ halo component,  which was not included in the treatment of the GRAVITY data in \cite{Wojtczak2023}. Neglecting the halo when deriving the observational results would potentially lead to an overestimation  of the size and underestimation of the photocentre shift magnitudes, depending on the flux ratio between resolved and unresolved region. If that is indeed the case, it may be possible to attain a similarly good fit to the remodelled data with a smaller magnetosphere that respects the co-rotation criterion.


\section*{Acknowledgements}

This work was supported by the “Programme National de Physique Stellaire” (PNPS) of CNRS/INSU co-funded by CEA and CNES. This project has received funding from the European Research Council (ERC) under the European Union’s Horizon 2020 research and innovation programme (grant agreement no. 742095; SPIDI: StarPlanets-Inner Disk-
Interactions, \hyperlink{http://www.spidi-eu.org}{http://www.spidi-eu.org}, and grant agreement no. 740651
NewWorlds). \linebreak
This work has made use of data from the European Space Agency
(ESA) mission Gaia (\hyperlink{https://www.cosmos.esa.int/gaia}{https://www.cosmos.esa.int/gaia}), processed by
the Gaia Data Processing and Analysis Consortium (DPAC, \hyperlink{https://www.
cosmos.esa.int/web/gaia/dpac/consortium}{https://www.
cosmos.esa.int/web/gaia/dpac/consortium}). Funding for the DPAC has
been provided by national institutions, in particular the institutions participating in the Gaia Multilateral Agreement. 
This research has made use of the Jean-Marie Mariotti Center \texttt{Aspro} service \footnote{Available at http://www.jmmc.fr/aspro}
V.G was supported for this research through a stipend from the International Max Planck Research School (IMPRS) for Astronomy and Astrophysics at the Universities of Bonn and Cologne, and from the Bonn-Cologne Graduate School of Physics and Astronomy (BCGS). 
A.C.G.has been supported by PRIN-INAF-MAIN-STREAM 2017 “Protoplanetary disks seen through the eyes of new-generation
instruments” and by PRIN-INAF 2019 “Spectroscopically tracing the disk dispersal evolution (STRADE)”
A.A.  and P.G. acknowledge supported by Fundação para a Ciência e a Tecnolo-
gia, with grants reference UIDB/00099/2020, SFRH/BSAB/142940/2018 and
PTDC/FIS-AST/7002/2020.
R.G.L. acknowledges support by Science Foundation Ireland under Grant
No. 18/SIRG/5597. 
\bibliographystyle{aa}
\bibliography{IDMRULup} 

\begin{thebibliography}{66}
\expandafter\ifx\csname natexlab\endcsname\relax\def\natexlab#1{#1}\fi

\bibitem[{{Alcal{\'a}} {et~al.}(2017){Alcal{\'a}}, {Manara}, {Natta}, {Frasca}, {Testi}, {Nisini}, {Stelzer}, {Williams}, {Antoniucci}, {Biazzo}, {Covino}, {Esposito}, {Getman}, \& {Rigliaco}}]{Alcala2017}
{Alcal{\'a}}, J.~M., {Manara}, C.~F., {Natta}, A., {et~al.} 2017, \aap, 600, A20

\bibitem[{{Alcal{\'a}} {et~al.}(2014){Alcal{\'a}}, {Natta}, {Manara}, {Spezzi}, {Stelzer}, {Frasca}, {Biazzo}, {Covino}, {Randich}, {Rigliaco}, {Testi}, {Comer{\'o}n}, {Cupani}, \& {D'Elia}}]{Alcala2014}
{Alcal{\'a}}, J.~M., {Natta}, A., {Manara}, C.~F., {et~al.} 2014, \aap, 561, A2

\bibitem[{{Alexander} {et~al.}(2014){Alexander}, {Pascucci}, {Andrews}, {Armitage}, \& {Cieza}}]{Alexander2014}
{Alexander}, R., {Pascucci}, I., {Andrews}, S., {Armitage}, P., \& {Cieza}, L. 2014, in Protostars and Planets VI, ed. H.~{Beuther}, R.~S. {Klessen}, C.~P. {Dullemond}, \& T.~{Henning}, 475--496

\bibitem[{{Berger} \& {Segransan}(2007)}]{Berger2007}
{Berger}, J.~P. \& {Segransan}, D. 2007, \nar, 51, 576

\bibitem[{{Blandford} \& {Payne}(1982)}]{BP1982}
{Blandford}, R.~D. \& {Payne}, D.~G. 1982, \mnras, 199, 883

\bibitem[{{Blinova} {et~al.}(2016){Blinova}, {Romanova}, \& {Lovelace}}]{Blinova2016}
{Blinova}, A.~A., {Romanova}, M.~M., \& {Lovelace}, R.~V.~E. 2016, \mnras, 459, 2354

\bibitem[{{Bouvier} {et~al.}(2007){Bouvier}, {Alencar}, {Harries}, {Johns-Krull}, \& {Romanova}}]{Bouvier2007}
{Bouvier}, J., {Alencar}, S.~H.~P., {Harries}, T.~J., {Johns-Krull}, C.~M., \& {Romanova}, M.~M. 2007, in Protostars and Planets V, ed. B.~{Reipurth}, D.~{Jewitt}, \& K.~{Keil}, 479

\bibitem[{{Bouvier} {et~al.}(2020){Bouvier}, {Perraut}, {Le Bouquin}, {Duvert}, {Dougados}, {Brandner}, {Benisty}, {Berger}, \& {Al{\'e}cian}}]{Bouvier2020a}
{Bouvier}, J., {Perraut}, K., {Le Bouquin}, J.~B., {et~al.} 2020, \aap, 636, A108

\bibitem[{{Donati} {et~al.}(2007){Donati}, {Jardine}, {Gregory}, {Petit}, {Bouvier}, {Dougados}, {M{\'e}nard}, {Collier Cameron}, {Harries}, {Jeffers}, \& {Paletou}}]{Donati2007}
{Donati}, J.~F., {Jardine}, M.~M., {Gregory}, S.~G., {et~al.} 2007, \mnras, 380, 1297

\bibitem[{{Drissen} {et~al.}(1989){Drissen}, {Bastien}, \& {St. -Louis}}]{Drissen1989}
{Drissen}, L., {Bastien}, P., \& {St. -Louis}, N. 1989, \aj, 97, 814

\bibitem[{{Eisner} {et~al.}(2010){Eisner}, {Monnier}, {Woillez}, {Akeson}, {Millan-Gabet}, {Graham}, {Hillenbrand}, {Pott}, {Ragland}, \& {Wizinowich}}]{Eisner2010}
{Eisner}, J.~A., {Monnier}, J.~D., {Woillez}, J., {et~al.} 2010, \apj, 718, 774

\bibitem[{{Fang} {et~al.}(2018){Fang}, {Pascucci}, {Edwards}, {Gorti}, {Banzatti}, {Flock}, {Hartigan}, {Herczeg}, \& {Dupree}}]{Fang2018}
{Fang}, M., {Pascucci}, I., {Edwards}, S., {et~al.} 2018, \apj, 868, 28

\bibitem[{{Ferreira} {et~al.}(2006){Ferreira}, {Dougados}, \& {Cabrit}}]{Ferreira2006}
{Ferreira}, J., {Dougados}, C., \& {Cabrit}, S. 2006, \aap, 453, 785

\bibitem[{{Ferreira} {et~al.}(2000){Ferreira}, {Pelletier}, \& {Appl}}]{Ferreira2000}
{Ferreira}, J., {Pelletier}, G., \& {Appl}, S. 2000, \mnras, 312, 387

\bibitem[{{Frasca} {et~al.}(2017){Frasca}, {Biazzo}, {Alcal{\'a}}, {Manara}, {Stelzer}, {Covino}, \& {Antoniucci}}]{Frasca2017}
{Frasca}, A., {Biazzo}, K., {Alcal{\'a}}, J.~M., {et~al.} 2017, \aap, 602, A33

\bibitem[{{Gaia Collaboration}(2020)}]{GaiaDR3}
{Gaia Collaboration}. 2020, VizieR Online Data Catalog, I/350

\bibitem[{{Garcia} {et~al.}(2001){Garcia}, {Ferreira}, {Cabrit}, \& {Binette}}]{Garcia2001}
{Garcia}, P.~J.~V., {Ferreira}, J., {Cabrit}, S., \& {Binette}, L. 2001, \aap, 377, 589

\bibitem[{{Gravity Collaboration} {et~al.}(2021{\natexlab{a}}){Gravity Collaboration}, {Ganci}, {Labadie}, {Klarmann}, {de Valon}, {Perraut}, {Benisty}, {Brandner}, {Caratti O Garatti}, {Dougados}, {Eupen}, {Garcia Lopez}, {Grellmann}, {Sanchez-Bermudez}, {Wojtczak}, {Garcia}, {Amorim}, {Baub{\"o}ck}, {Berger}, {Caselli}, {Cl{\'e}net}, {Coud{\'e} Du Foresto}, {de Zeeuw}, {Drescher}, {Duvert}, {Eckart}, {Eisenhauer}, {Filho}, {Gao}, {Gendron}, {Genzel}, {Gillessen}, {Heissel}, {Henning}, {Hippler}, {Horrobin}, {Hubert}, {Jim{\'e}nez-Rosales}, {Jocou}, {Kervella}, {Lacour}, {Lapeyr{\`e}re}, {Le Bouquin}, {L{\'e}na}, {Ott}, {Paumard}, {Perrin}, {Pfuhl}, {Hei{\ss}el}, {Rousset}, {Scheithauer}, {Shangguan}, {Shimizu}, {Stadler}, {Straub}, {Straubmeier}, {Sturm}, {van Dishoeck}, {Vincent}, {von Fellenberg}, {Widmann}, \& {Woillez}}]{Ganci2021}
{Gravity Collaboration}, {Ganci}, V., {Labadie}, L., {et~al.} 2021{\natexlab{a}}, \aap, 655, A112

\bibitem[{{Gravity Collaboration} {et~al.}(2020){Gravity Collaboration}, {Garcia Lopez}, {Natta}, {Caratti o Garatti}, {Ray}, {Fedriani}, {Koutoulaki}, {Klarmann}, {Perraut}, {Sanchez-Bermudez}, {Benisty}, {Dougados}, {Labadie}, {Brandner}, {Garcia}, {Henning}, {Caselli}, {Duvert}, {de Zeeuw}, {Grellmann}, {Abuter}, {Amorim}, {Baub{\"o}ck}, {Berger}, {Bonnet}, {Buron}, {Cl{\'e}net}, {Coud{\'e} Du Foresto}, {de Wit}, {Eckart}, {Eisenhauer}, {Filho}, {Gao}, {Garcia Dabo}, {Gendron}, {Genzel}, {Gillessen}, {Habibi}, {Haubois}, {Haussmann}, {Hippler}, {Hubert}, {Horrobin}, {Jimenez Rosales}, {Jocou}, {Kervella}, {Kolb}, {Lacour}, {Le Bouquin}, {L{\'e}na}, {Ott}, {Paumard}, {Perrin}, {Pfuhl}, {Ramirez}, {Rau}, {Rousset}, {Scheithauer}, {Shangguan}, {Stadler}, {Straub}, {Straubmeier}, {Sturm}, {van Dishoeck}, {Vincent}, {von Fellenberg}, {Widmann}, {Wieprecht}, {Wiest}, {Wiezorrek}, {Woillez}, {Yazici}, \& {Zins}}]{GarciaLopez2020}
{Gravity Collaboration}, {Garcia Lopez}, R., {Natta}, A., {et~al.} 2020, Nature, 584, 547

\bibitem[{{Gravity Collaboration} {et~al.}(2021{\natexlab{b}}){Gravity Collaboration}, {Perraut}, {Labadie}, {Bouvier}, {M{\'e}nard}, {Klarmann}, {Dougados}, {Benisty}, {Berger}, {Bouarour}, {Brandner}, {Caratti O Garatti}, {Caselli}, {de Zeeuw}, {Garcia-Lopez}, {Henning}, {Sanchez-Bermudez}, {Sousa}, {van Dishoeck}, {Al{\'e}cian}, {Amorim}, {Cl{\'e}net}, {Davies}, {Drescher}, {Duvert}, {Eckart}, {Eisenhauer}, {F{\"o}rster-Schreiber}, {Garcia}, {Gendron}, {Genzel}, {Gillessen}, {Grellmann}, {Hei{\ss}el}, {Hippler}, {Horrobin}, {Hubert}, {Jocou}, {Kervella}, {Lacour}, {Lapeyr{\`e}re}, {Le Bouquin}, {L{\'e}na}, {Lutz}, {Ott}, {Paumard}, {Perrin}, {Scheithauer}, {Shangguan}, {Shimizu}, {Stadler}, {Straub}, {Straubmeier}, {Sturm}, {Tacconi}, {Vincent}, {von Fellenberg}, \& {Widmann}}]{Perraut2021}
{Gravity Collaboration}, {Perraut}, K., {Labadie}, L., {et~al.} 2021{\natexlab{b}}, \aap, 655, A73

\bibitem[{{Gravity Collaboration} {et~al.}(2023{\natexlab{a}}){Gravity Collaboration}, {Soulain}, {Perraut}, {Bouvier}, {Pantolmos}, {Caratti O Garatti}, {Caselli}, {Garcia}, {Lopez}, {Aimar}, {Amorin}, {Benisty}, {Berger}, {Bourdarot}, {Brandner}, {Cl{\'e}net}, {de Zeeuw}, {Davies}, {Drescher}, {Eckart}, {Eisenhauer}, {Schreiber}, {Gendron}, {Genzuel}, {Gillessen}, {Hei{\ss}el}, {Henning}, {Hippler}, {Horrobin}, {Jocou}, {Kervella}, {Labadie}, {Lacour}, {Lapeyrere}, {Le Bouquin}, {L{\'e}na}, {Lutz}, {Mang}, {Ott}, {Paumard}, {Perrin}, {Sanchez}, {Scheithauer}, {Shangguan}, {Shimizu}, {Straub}, {Straubmeier}, {Sturm}, {Tacconi}, {Vincent}, {van Dishoeck}, {Widmann}, {Wieprecht}, {Wiezorrek}, \& {Yazici}}]{Soulain2023}
{Gravity Collaboration}, {Soulain}, A., {Perraut}, K., {et~al.} 2023{\natexlab{a}}, \aap, 674, A203

\bibitem[{{Gravity Collaboration} {et~al.}(2023{\natexlab{b}}){Gravity Collaboration}, {Wojtczak}, {Labadie}, {Perraut}, {Tessore}, {Soulain}, {Ganci}, {Bouvier}, {Dougados}, {Al{\'e}cian}, {Nowacki}, {Cozzo}, {Brandner}, {Caratti O Garatti}, {Garcia}, {Garcia Lopez}, {Sanchez-Bermudez}, {Amorim}, {Benisty}, {Berger}, {Bourdarot}, {Caselli}, {Cl{\'e}net}, {de Zeeuw}, {Davies}, {Drescher}, {Duvert}, {Eckart}, {Eisenhauer}, {Eupen}, {F{\"o}rster-Schreiber}, {Gendron}, {Gillessen}, {Grant}, {Grellmann}, {Hei{\ss}el}, {Henning}, {Hippler}, {Horrobin}, {Hubert}, {Jocou}, {Kervella}, {Lacour}, {Lapeyr{\`e}re}, {Le Bouquin}, {L{\'e}na}, {Lutz}, {Mang}, {Ott}, {Paumard}, {Perrin}, {Scheithauer}, {Shangguan}, {Shimizu}, {Spezzano}, {Straub}, {Straubmeier}, {Sturm}, {van Dishoeck}, {Vincent}, \& {Widmann}}]{Wojtczak2023}
{Gravity Collaboration}, {Wojtczak}, J.~A., {Labadie}, L., {et~al.} 2023{\natexlab{b}}, \aap, 669, A59

\bibitem[{{Hartmann} {et~al.}(1982){Hartmann}, {Avrett}, \& {Edwards}}]{Hartmann1982}
{Hartmann}, L., {Avrett}, E., \& {Edwards}, S. 1982, \apj, 261, 279

\bibitem[{{Hartmann} {et~al.}(2016){Hartmann}, {Herczeg}, \& {Calvet}}]{Hartmann2016}
{Hartmann}, L., {Herczeg}, G., \& {Calvet}, N. 2016, \araa, 54, 135

\bibitem[{{Hartmann} {et~al.}(1994){Hartmann}, {Hewett}, \& {Calvet}}]{Hartmann1994}
{Hartmann}, L., {Hewett}, R., \& {Calvet}, N. 1994, \apj, 426, 669

\bibitem[{{Hartmann} \& {Stauffer}(1989)}]{HartmannStauffer1989}
{Hartmann}, L. \& {Stauffer}, J.~R. 1989, \aj, 97, 873

\bibitem[{{Hoffmeister}(1965)}]{Hoffmeister1965}
{Hoffmeister}, C. 1965, Veroeffentlichungen der Sternwarte Sonneberg, 6, 123

\bibitem[{{Huang} {et~al.}(2018){Huang}, {Andrews}, {Dullemond}, {Isella}, {P{\'e}rez}, {Guzm{\'a}n}, {{\"O}berg}, {Zhu}, {Zhang}, {Bai}, {Benisty}, {Birnstiel}, {Carpenter}, {Hughes}, {Ricci}, {Weaver}, \& {Wilner}}]{Huang2018}
{Huang}, J., {Andrews}, S.~M., {Dullemond}, C.~P., {et~al.} 2018, \apjl, 869, L42

\bibitem[{{Johnstone} {et~al.}(2014){Johnstone}, {Jardine}, {Gregory}, {Donati}, \& {Hussain}}]{Johnstone2014}
{Johnstone}, C.~P., {Jardine}, M., {Gregory}, S.~G., {Donati}, J.~F., \& {Hussain}, G. 2014, \mnras, 437, 3202

\bibitem[{{Knigge} {et~al.}(1995){Knigge}, {Woods}, \& {Drew}}]{Knigge1995}
{Knigge}, C., {Woods}, J.~A., \& {Drew}, J.~E. 1995, \mnras, 273, 225

\bibitem[{{Kurosawa} {et~al.}(2006){Kurosawa}, {Harries}, \& {Symington}}]{Kurosawa2006}
{Kurosawa}, R., {Harries}, T.~J., \& {Symington}, N.~H. 2006, \mnras, 370, 580

\bibitem[{{Kurosawa} {et~al.}(2016){Kurosawa}, {Kreplin}, {Weigelt}, {Natta}, {Benisty}, {Isella}, {Tatulli}, {Massi}, {Testi}, {Kraus}, {Duvert}, {Petrov}, \& {Stee}}]{Kurosawa2016}
{Kurosawa}, R., {Kreplin}, A., {Weigelt}, G., {et~al.} 2016, \mnras, 457, 2236

\bibitem[{{Kurosawa} {et~al.}(2011){Kurosawa}, {Romanova}, \& {Harries}}]{Kurosawa2011}
{Kurosawa}, R., {Romanova}, M.~M., \& {Harries}, T.~J. 2011, \mnras, 416, 2623

\bibitem[{{Le Bouquin} {et~al.}(2009){Le Bouquin}, {Absil}, {Benisty}, {Massi}, {M{\'e}rand}, \& {Stefl}}]{LeBouquin2009}
{Le Bouquin}, J.~B., {Absil}, O., {Benisty}, M., {et~al.} 2009, \aap, 498, L41

\bibitem[{{Lima} {et~al.}(2010){Lima}, {Alencar}, {Calvet}, {Hartmann}, \& {Muzerolle}}]{Lima2010}
{Lima}, G.~H.~R.~A., {Alencar}, S.~H.~P., {Calvet}, N., {Hartmann}, L., \& {Muzerolle}, J. 2010, \aap, 522, A104

\bibitem[{{Mahdavi} \& {Kenyon}(1998)}]{Mahdavi1998}
{Mahdavi}, A. \& {Kenyon}, S.~J. 1998, \apj, 497, 342

\bibitem[{{Matt} \& {Pudritz}(2005)}]{MattPudritz2005}
{Matt}, S. \& {Pudritz}, R.~E. 2005, \apjl, 632, L135

\bibitem[{{McGinnis} {et~al.}(2020){McGinnis}, {Bouvier}, \& {Gallet}}]{McGinnis2020}
{McGinnis}, P., {Bouvier}, J., \& {Gallet}, F. 2020, \mnras, 497, 2142

\bibitem[{{Monnier} \& {Millan-Gabet}(2002)}]{Monnier2002}
{Monnier}, J.~D. \& {Millan-Gabet}, R. 2002, \apj, 579, 694

\bibitem[{{Muzerolle} {et~al.}(1998){Muzerolle}, {Calvet}, \& {Hartmann}}]{Muzerolle1998}
{Muzerolle}, J., {Calvet}, N., \& {Hartmann}, L. 1998, \apj, 492, 743

\bibitem[{{Muzerolle} {et~al.}(2001){Muzerolle}, {Calvet}, \& {Hartmann}}]{Muzerolle2001}
{Muzerolle}, J., {Calvet}, N., \& {Hartmann}, L. 2001, \apj, 550, 944

\bibitem[{{Muzerolle} {et~al.}(2004){Muzerolle}, {D'Alessio}, {Calvet}, \& {Hartmann}}]{Muzerolle2004}
{Muzerolle}, J., {D'Alessio}, P., {Calvet}, N., \& {Hartmann}, L. 2004, \apj, 617, 406

\bibitem[{{Natta} {et~al.}(2001){Natta}, {Prusti}, {Neri}, {Wooden}, {Grinin}, \& {Mannings}}]{Natta2001}
{Natta}, A., {Prusti}, T., {Neri}, R., {et~al.} 2001, \aap, 371, 186

\bibitem[{{Pantolmos} {et~al.}(2020){Pantolmos}, {Zanni}, \& {Bouvier}}]{PantolmosZanni2020}
{Pantolmos}, G., {Zanni}, C., \& {Bouvier}, J. 2020, \aap, 643, A129

\bibitem[{{Pinte} {et~al.}(2009){Pinte}, {Harries}, {Min}, {Watson}, {Dullemond}, {Woitke}, {M{\'e}nard}, \& {Dur{\'a}n-Rojas}}]{Pinte2009}
{Pinte}, C., {Harries}, T.~J., {Min}, M., {et~al.} 2009, \aap, 498, 967

\bibitem[{{Pinte} {et~al.}(2006){Pinte}, {M{\'e}nard}, {Duch{\^e}ne}, \& {Bastien}}]{Pinte2006}
{Pinte}, C., {M{\'e}nard}, F., {Duch{\^e}ne}, G., \& {Bastien}, P. 2006, \aap, 459, 797

\bibitem[{{Rebull} {et~al.}(2020){Rebull}, {Stauffer}, {Cody}, {Hillenbrand}, {Bouvier}, {Roggero}, \& {David}}]{Rebull2020}
{Rebull}, L.~M., {Stauffer}, J.~R., {Cody}, A.~M., {et~al.} 2020, \aj, 159, 273

\bibitem[{{Romanova} {et~al.}(2008){Romanova}, {Kulkarni}, \& {Lovelace}}]{Romanova2008-2}
{Romanova}, M.~M., {Kulkarni}, A.~K., \& {Lovelace}, R. V.~E. 2008, \apjl, 673, L171

\bibitem[{{Romanova} \& {Owocki}(2015)}]{Romanova2015}
{Romanova}, M.~M. \& {Owocki}, S.~P. 2015, \ssr, 191, 339

\bibitem[{{Romanova} \& {Owocki}(2016)}]{Romanova2016}
{Romanova}, M.~M. \& {Owocki}, S.~P. 2016, in The Strongest Magnetic Fields in the Universe: Space Sciences Series of ISSI, Vol.~54, 347

\bibitem[{{Setterholm} {et~al.}(2018){Setterholm}, {Monnier}, {Davies}, {Kreplin}, {Kraus}, {Baron}, {Aarnio}, {Berger}, {Calvet}, {Cur{\'e}}, {Kanaan}, {Kloppenborg}, {Le Bouquin}, {Millan-Gabet}, {Rubinstein}, {Sitko}, {Sturmann}, {ten Brummelaar}, \& {Touhami}}]{Setterholm2018}
{Setterholm}, B.~R., {Monnier}, J.~D., {Davies}, C.~L., {et~al.} 2018, \apj, 869, 164

\bibitem[{{Shu} {et~al.}(1994){Shu}, {Najita}, {Ostriker}, {Wilkin}, {Ruden}, \& {Lizano}}]{Shu1994}
{Shu}, F., {Najita}, J., {Ostriker}, E., {et~al.} 1994, \apj, 429

\bibitem[{{Siwak} {et~al.}(2016){Siwak}, {Ogloza}, {Rucinski}, {Moffat}, {Matthews}, {Cameron}, {Guenther}, {Kuschnig}, {Rowe}, {Sasselov}, \& {Weiss}}]{Siwak2016}
{Siwak}, M., {Ogloza}, W., {Rucinski}, S.~M., {et~al.} 2016, \mnras, 456, 3972

\bibitem[{{Stempels} {et~al.}(2007){Stempels}, {Gahm}, \& {Petrov}}]{Stempels2007}
{Stempels}, H.~C., {Gahm}, G.~F., \& {Petrov}, P.~P. 2007, \aap, 461, 253

\bibitem[{{Stock} {et~al.}(2022){Stock}, {McGinnis}, {Caratti o Garatti}, {Natta}, \& {Ray}}]{Stock22}
{Stock}, C., {McGinnis}, P., {Caratti o Garatti}, A., {Natta}, A., \& {Ray}, T.~P. 2022, \aap, 668, A94

\bibitem[{{Tabone} {et~al.}(2022){Tabone}, {Rosotti}, {Cridland}, {Armitage}, \& {Lodato}}]{Tabone22}
{Tabone}, B., {Rosotti}, G.~P., {Cridland}, A.~J., {Armitage}, P.~J., \& {Lodato}, G. 2022, \mnras, 512, 2290

\bibitem[{{Takami} {et~al.}(2003){Takami}, {Bailey}, \& {Chrysostomou}}]{Takami2003}
{Takami}, M., {Bailey}, J., \& {Chrysostomou}, A. 2003, \aap, 397, 675

\bibitem[{{Takasao} {et~al.}(2022){Takasao}, {Tomida}, {Iwasaki}, \& {Suzuki}}]{Takasao2022}
{Takasao}, S., {Tomida}, K., {Iwasaki}, K., \& {Suzuki}, T.~K. 2022, \apj, 941, 73

\bibitem[{{Tessore} {et~al.}(2021){Tessore}, {Pinte}, {Bouvier}, \& {M{\'e}nard}}]{Tessore21}
{Tessore}, B., {Pinte}, C., {Bouvier}, J., \& {M{\'e}nard}, F. 2021, \aap, 647, A27

\bibitem[{{Tessore} {et~al.}(2023){Tessore}, {Soulain}, {Pantolmos}, {Bouvier}, {Pinte}, \& {Perraut}}]{Tessore23}
{Tessore}, B., {Soulain}, A., {Pantolmos}, G., {et~al.} 2023, \aap, 671, A129

\bibitem[{{Watson} {et~al.}(2016){Watson}, {Calvet}, {Fischer}, {Forrest}, {Manoj}, {Megeath}, {Melnick}, {Najita}, {Neufeld}, {Sheehan}, {Stutz}, \& {Tobin}}]{Watson2016}
{Watson}, D.~M., {Calvet}, N.~P., {Fischer}, W.~J., {et~al.} 2016, \apj, 828, 52

\bibitem[{{Weber} {et~al.}(2020){Weber}, {Ercolano}, {Picogna}, {Hartmann}, \& {Rodenkirch}}]{Weber2020}
{Weber}, M.~L., {Ercolano}, B., {Picogna}, G., {Hartmann}, L., \& {Rodenkirch}, P.~J. 2020, \mnras, 496, 223

\bibitem[{{Weigelt} {et~al.}(2011){Weigelt}, {Grinin}, {Groh}, {Hofmann}, {Kraus}, {Miroshnichenko}, {Schertl}, {Tambovtseva}, {Benisty}, {Driebe}, {Lagarde}, {Malbet}, {Meilland}, {Petrov}, \& {Tatulli}}]{Weigelt2011}
{Weigelt}, G., {Grinin}, V.~P., {Groh}, J.~H., {et~al.} 2011, \aap, 527, A103

\bibitem[{{Whelan} {et~al.}(2021){Whelan}, {Pascucci}, {Gorti}, {Edwards}, {Alexander}, {Sterzik}, \& {Melo}}]{Whelan2021}
{Whelan}, E.~T., {Pascucci}, I., {Gorti}, U., {et~al.} 2021, \apj, 913, 43

\bibitem[{{Wilson} {et~al.}(2022){Wilson}, {Matt}, {Harries}, \& {Herczeg}}]{Wilson2022}
{Wilson}, T.~J.~G., {Matt}, S., {Harries}, T.~J., \& {Herczeg}, G.~J. 2022, \mnras, 514, 2162

\bibitem[{{Zanni} \& {Ferreira}(2013)}]{Zanni2013}
{Zanni}, C. \& {Ferreira}, J. 2013, \aap, 550, A99

\end{thebibliography}

\appendix
\section{Parameter dependencies of the models}
\label{sec:sensitivity}
\begin{table*}[h]
\footnotesize
\caption{Magnetospheric accretion model response properties.}
\centering
\begin{tabular}{ccccccccccc}
	&	P$_r$		&	$\Delta$P 	&	$\Delta$Flux 	&	$\Delta$Size 	&	$\Delta$PC	&	Mean $\Delta$ 	&	$\xi_{Flux}$	&	$\xi_{Size}$	&	$\xi_{PC}$	&	$\xi_{Tot}$	\\
 &  & [\%] & [\%] & [\%] & [\%] & [\%] & [\%] &[\%] &[\%] &[\%] \\ \hline \hline
T$_{mag}$	&	8600 K	&	16	&	22	&	27	&	73	&	41	&	134	&	167	&	448	&	250	\\[0.1cm]
R$_*$	&	2.5 R$_{\odot}$	&	24	&	5	&	21	&	27	&	17	&	22	&	86	&	111	&	73	\\[0.1cm]
R$_{mi}$	&	7 R$_*$	&	43	&	10	&	29	&	39	&	26	&	23	&	69	&	90	&	61	\\[0.1cm]
P$_*$	&	7 days	&	45	&	3	&	6	&	38	&	16	&	7	&	14	&	84	&	35	\\[0.1cm]
$\dot{M}_{acc}$	&	23 $\times 10^{-8}$ M$_{\odot}$ yr$^{-1}$	&	22	&	8	&	4	&	11	&	8	&	35	&	16	&	52	&	35	\\[0.1cm]
$\delta R_{m}$	&	2 R$_*$	&	50	&	12	&	4	&	26	&	14	&	24	&	8	&	51	&	28	\\[0.1cm]
M$_*$	&	0.8 M$_{\odot}$	&	38	&	7	&	4	&	12	&	8	&	20	&	10	&	32	&	21	\\[0.1cm]
T$_*$	&	4050 K	&	2	&	0	&	0	&	0	&	0	&	2	&	0	&	4	&	2	\\[0.1cm] \hline
Average	&			&	30	&	8	&	12	&	28	&	16	&	33	&	46	&	109	&	63	\\
\hline \hline
\end{tabular}   
\tablefoot{All quantities marked with $\Delta$ are mean parameter variations of the magnetospheric accretion model, given in relation to the reference parameter P$_r$. $\xi$ denotes the resulting responses of the computed synthetic observables (line-to-continuum flux ratio, characteristic emission region size, and emission region offset), as computed according to Eq. \ref{eq:response}.}
\end{table*}

While the parameter descriptions laid out in sections \ref{sec:Macc} and \ref{sec:dwmodel}, and their associated equations given in \cite{Hartmann1982} and \cite{Knigge1995}, may outline the general impact of a parameter change on the densities and velocities in the system, they cannot straightforwardly connect the effect of such a variation with the resulting emission line profiles and possibly even less so with the interferometric quantities. Due to the complex interplay between atomic populations, radiative transitions, and the spatial distribution of velocities and densities, it is more instructive to directly analyse the effective change in line strength, emission region size, and photocentre distribution when increasing or decreasing a single parameter by a certain margin. In this manner, both the total sensitivity of the model as well as the partial sensitivities of these individual quantities to changes in a specific parameter may be constrained and compared. The limitations to this approach lie predominantly in the need to compute a new model for each parameter change and the time investment required to do so. Given the relatively large amount of parameters that we consider, in principle, fit for variation up to a point, we limit this analysis to one increase and one decrease relative to the parameter value of an arbitrarily chosen reference model. The step size of the variation is not set consistently but rather chosen based on relevant upper and lower limits from the literature for parameters which have known observational constraints, such as the stellar radius or mass. 
Other parameters, for which such constraints are not available, are varied to a degree that results in a similar change in the line-to-continuum flux ratio. While a purely theoretical study of the models would certainly look at a broader range of parameter changes, we remind that the purpose of this analysis was primarily to determine which quantities were the most relevant to the interpretation of specific observational results. \\ \indent
We differentiate between individual parameters based on their influence on the results. To this end, we present multiple ways to quantify the impact. First, we consider the relative change $\Delta$O as a percentage of the respective reference model parameter:
\begin{align} 
    \Delta O = \left \lvert \frac{O_v - O_{r}}{O_{r}} \right \rvert.
\label{eq:deltay}
\end{align}
Here, O is one of the observables we compute from the model images, i.e. line-to-continuum flux ratio, characteristic size of the emission region or photocentre offset of the emission region. The subscript v designates that the observable belongs to the variation model, while r designates the reference model. We choose to employ a relative measure of the change primarily due to the very different scales between these three observables. If we were to consider the absolute change, the large flux ratio values would dominate the average over the small sizes and photocentre shifts. In particular, we consider here the mean relative change between the parameter increase and the parameter decrease.
We further compute the unweighted average of this metric over all considered wavelengths across the \bg \ feature. Then, we subsequently set this value in relation to the relative parameter step $\Delta$P, which is computed analogously according to Eq. \ref{eq:deltay} for a variation model parameter P$_v$ and a reference model parameter P$_r$. Thus, we arrive at a measure for the relative mean observable change, adjusted by the parameter step size. We refer to this quantity as the model observable O response $\xi_{O}$ to a parameter P change $\Delta$P.:
\begin{align}\label{eq:response}
    \xi_O=\frac{\Delta O}{\Delta P}.
\end{align} 

\subsection{Magnetospheric accretion }


\begin{table*}[!t] 
\footnotesize
\caption{Disk wind model response properties.}
\label{tab:WPV}
\centering
\begin{tabular}{ccccccccccc}
	&	P$_r$		&	$\Delta$ P 		&	$\Delta$ Flux 	&	$\Delta$ Size 	&	$\Delta$ PC 	&	Mean $\Delta$ 	&	$\xi_{Flux}$	&	$\xi_{Size}$	&	$\xi_{PC}$	&	$\xi_{Tot}$	\\
 &  & [\%] & [\%] & [\%] & [\%] & [\%] & [\%] &[\%] &[\%] &[\%] \\ \hline \hline
z$_{crit}$	&	0.0516 au	&	3		&	15	&	14	&	35	&	21	&	522	&	473	&	1195	&	730		\\[0.1cm]
$\beta_{wind}$	&	0.62		&	6		&	27	&	21	&	63	&	37	&	487	&	376	&	1114	&	659		\\[0.1cm]
$\dot{M}_{loss}$	&	5.1 $\times 10^{-8}$ M$_{\odot}$ yr$^{-1}$	&	4		&	18	&	15	&	36	&	23	&	465	&	375	&	922	&	588		\\[0.1cm]
$\delta R_{w}$	&	7.1 R$_*$	&	20		&	40	&	29	&	94	&	54	&	200	&	147	&	469	&	272		\\[0.1cm]
R$_{wi}$	&	5 R$_*$	&	14		&	15	&	11	&	69	&	31	&	104	&	75	&	489	&	223		\\[0.1cm]
d$_{fp}$	&	15 R$_*$	&	33		&	39	&	34	&	138	&	70	&	118	&	103	&	413	&	211		\\[0.1cm]
R$_*$	&	2.5 R$_{\odot}$	&	24		&	11	&	14	&	93	&	39	&	46	&	57	&	386	&	163		\\[0.1cm]
R$_{dd}$	&	14.5 R$_*$	&	34		&	14	&	12	&	141	&	56	&	40	&	36	&	408	&	161		\\[0.1cm]
M$_*$	&	0.8 M$_{\odot}$	&	38		&	41	&	29	&	88	&	53	&	109	&	78	&	234	&	140		\\[0.1cm]
$\alpha_{wind}$	&	0.4		&	25		&	26	&	18	&	47	&	31	&	104	&	72	&	190	&	122		\\[0.1cm]
T$_*$	&	4050 K	&	5		&	3	&	2	&	7	&	4	&	68	&	44	&	140	&	84		\\[0.1cm]
P$_*$	&	7 days	&	45		&	0	&	0	&	0	&	0	&	0	&	0	&	0	&	0		\\[0.1cm] \hline
Average	&			&	21		&	21	&	17	&	67	&	35	&	188	&	153	&	497	&	279		\\
\hline \hline
\end{tabular}   
\tablefoot{All quantities marked with $\Delta$ are mean parameter variations of the disk wind model, given in relation to the reference parameter P$_r$. $\xi$ denotes the resulting responses of the computed synthetic observables (line-to-continuum flux ratio, characteristic emission region size, and emission region offset), as computed according to Eq. \ref{eq:response}.}
\end{table*}

\begin{figure*}[t] 
    \centering
    \includegraphics[width=0.9\linewidth]{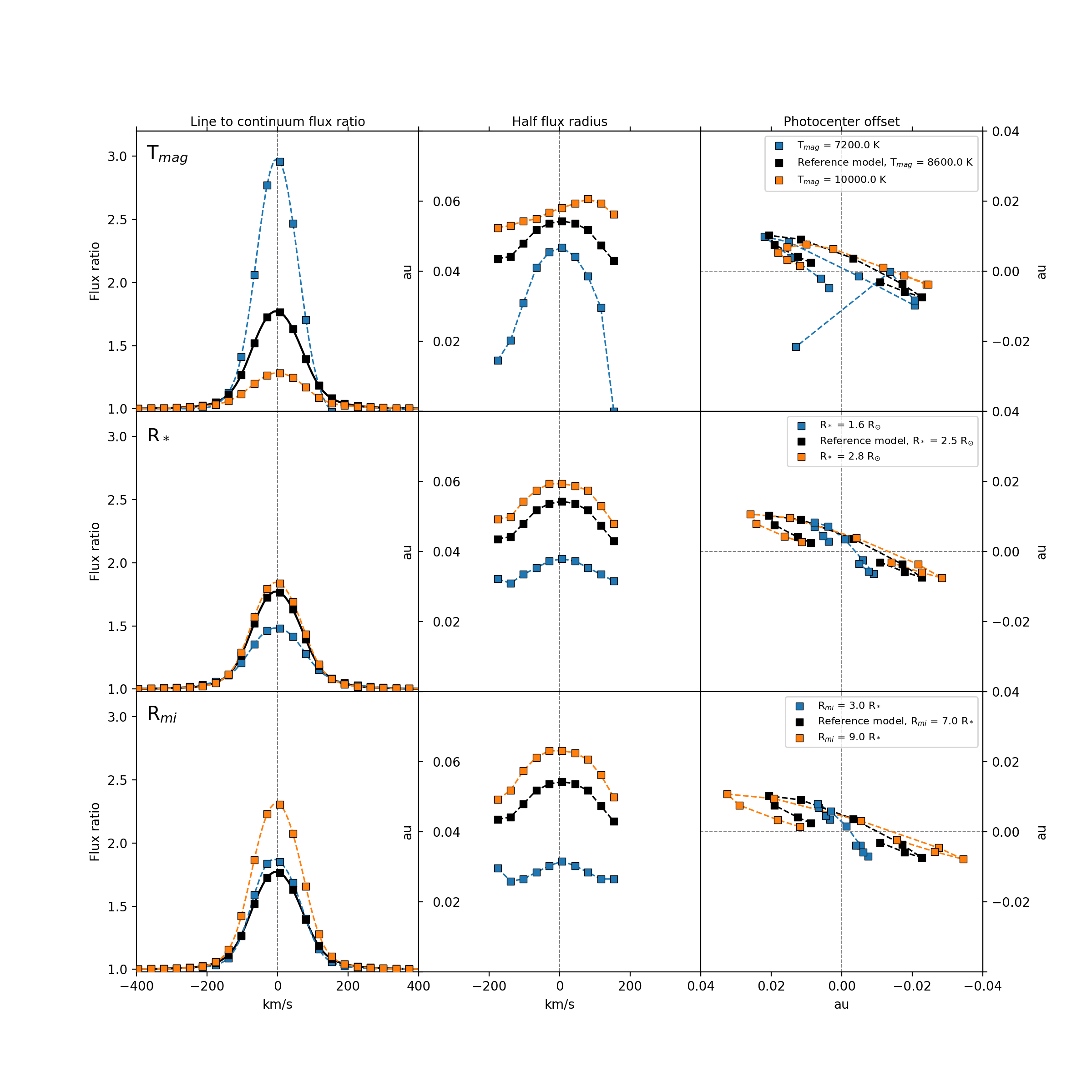}
    \caption{Observable dependencies on parameter changes for the top three \textbf{magnetospheric accretion} parameters as ranked in Table \ref{tab:MPV}. The left column depicts the changes in continuum normalised line flux. The centre column shows a characteristic interferometric size, obtained as the half width at half maximum (HWHM) of a geometric Gaussian disk model. The right column illustrates the spatial distribution of line emission photocentres at different velocity channels.
    }
    \label{fig:mpv}
\end{figure*}

In Table \ref{tab:MPV} we present the relative effects of a parameter change on the magnetospheric accretion model observables, as well as the response metrics computed based on Eq. \ref{eq:response}. When ranked by their respective $\xi_{Tot}$ values, the maximum temperature in the magnetospheric funnel flows T$_{mag}$, the stellar Radius R$_*$, and the inner magnetospheric truncation radius R$_{mi}$ come out top in terms of their overall impact on the resulting model observables.  Figure \ref{fig:mpv} shows the deviations of the line profile, the size, and photocentre shifts from the reference model for the across all channels of the \bg \ feature for these quantities. Here, the effect of a magnetospheric temperature variation on all three observables stands out as particularly noteworthy when taking into account the relatively small variation step width. The relative changes in photocentre shifts and sizes are on the order of factor 2 to 4 higher when compared even to the other two high impact parameters and on the order of factor 10 when compared to the mass accretion rate of the system, for which the variation step width of 22$\%$ is similar to T$_{mag}$ at 16$\%$. Figure \ref{fig:mpv} shows that an increase in magnetospheric temperature from the reference value of 8600 K leads to a decrease in continuum-normalised flux and simultaneously to an increase in the emission regions characteristic size. The former suggests that, at such high temperatures, a significant amount of continuum emission is produced in the accretion columns due to bound-free transitions as the hydrogen becomes increasingly ionised. The larger characteristic size stems from a flatter distribution of flux across the entire magnetosphere, which leads to a spatially more extended full width at half maximum of the Gaussian disk model used to fit the characteristic size. For the same reason, the photocentre appears slightly less extended, as the more evenly distributed emission throughout the very hot magnetosphere symmetrises the emission region relative to the reference model. \\ \indent 
Similar reasoning can be applied when breaking down the effects of other parameter changes, although in the case of the stellar radius one should take care to separate the physical influence of the radius from model scaling effects. Since we define the size of our truncation radius and magnetospheric widths in units of stellar radii, we are effectively changing the absolute physical size of the magnetosphere by varying the stellar radius and thus also the position of the field line anchor points as they appear in those equations. This in turn explains why the impact a stellar radius variation seems to beat an adjustment of the truncation radius, both in terms of $\xi_{Tot}$ and also in the partial metric for the characteristic size alone. The characteristic size of the system is a product of both R$_{mi}$ and, albeit to a significantly lesser degree, $\delta$R$_{m}$, which are both defined in units of R$_*$, whereas setting the truncation radius directly will only affect $R_{mi}$.  \\ \indent
By contrast, it appears that the stellar mass M$_*$ and photospheric temperature T$_*$ only marginally influence the outcome of the computations. This is particularly true for the photospheric temperature, which, even when accounting for the very small parameter variation of about 2\%, produces a negligible change in observables compared to the other quantities. The low impact of T$_*$ on the interferometric quantities is not surprising, given that the characteristic size of the emission region is based on the continuum extracted pure line visibilities and the pure line differential phase. The fact that some variation in photocentre shift with changing photospheric temperature can be observed at all is likely caused by the shifting weights between the stellar continuum and the magnetospheric continuum. Indeed, the former will produce a small non-zero continuum phase at higher inclinations due to the presence of the ring-like shock region close to the pole, while the magnetospheric continuum is centrosymmetric for an axisymmetric configuration, such as used here. \\ \indent
Finally, the stellar period does not affect the flux ratio or characteristic size to the same degree as R$_*$, but has a comparable effect on the photocentre shift, due to increased line broadening at higher rotational velocities of the star. Since the magnetospheric funnels rotate at the same velocity, observing the rotating gas stream at a non-zero inclination will lead to a broader line profile. The change in flux ratio caused by this is not necessarily large, as demonstrated by the low $\xi_{Flux}$ of only 7\%. The redistribution of flux to higher velocities, however, translates to a relatively large change in magnetospheric symmetries at those velocities and thus the more drastic variations in the photocentres per velocity component.

\subsection{Disk wind}

\begin{figure*}[t] 
    \centering
    \includegraphics[width=0.9\linewidth]{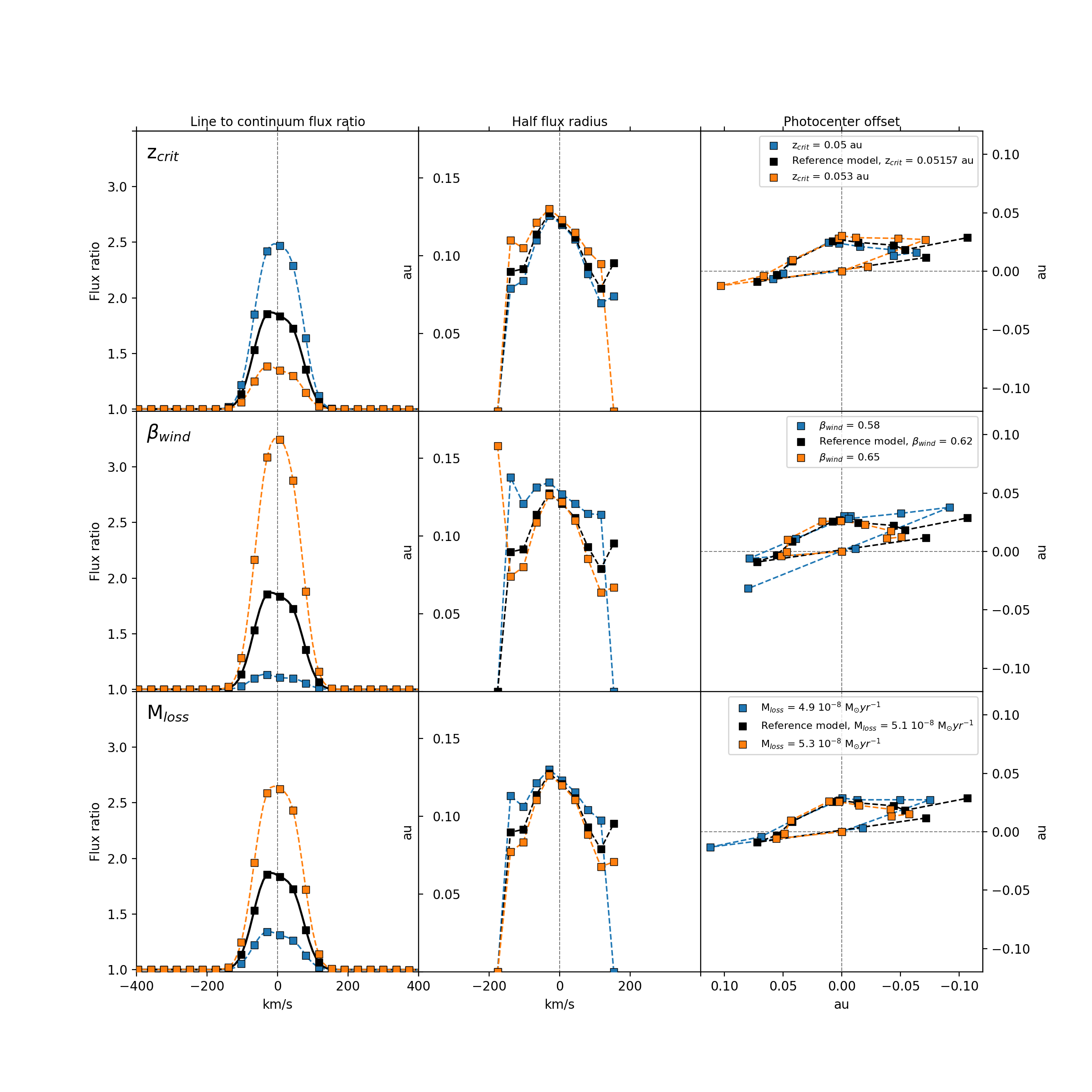}
    \caption{Observable dependencies on parameter changes for the top three \textbf{disk wind} parameters as ranked in Table \ref{tab:WPV}. The left column depicts the changes in continuum normalised line flux. The centre column shows a characteristic interferometric size, obtained as the half width at half maximum (HWHM) of a geometric Gaussian disk model. The right column illustrates the spatial distribution of line emission photocentres at different velocity channels.
    }
    \label{fig:wpv}
\end{figure*}

In Table \ref{tab:WPV} we present an analogous ranking of the response metric $\xi_{tot}$ for the disk wind model. It demonstrates that again three parameters perform significantly stronger compared to the rest in terms of their impact on the three observables. For the wind model, we find that the critical height z$_{crit}$, the velocity law parameter $\beta_{wind}$, and the global mass loss rate $\dot{M}_{loss}$ have far above average impact on the resulting line profiles, characteristic sizes, and photocentre shifts. The response metrics for these three parameters are broadly similar across the three observables. For each of them, the step width-weighted parameter variation causes a particularly large relative change in the photocentre profile, for which the $\xi_{PC}$ is about a factor two larger than the corresponding flux ratio metric, whereas the respective $\xi_{Size}$ values are similar to, but slightly smaller than, the flux ratio metric. In Figure \ref{fig:wpv} we show the corresponding changes in flux ratio, size, and photocentre offset. \\ \indent
It is obvious that a higher cutoff z$_{crit}$ of the wind leads to a weaker line-to-continuum ratio. This is expected, as the density along a stream line falls with increasing distance from the midplane. As a consequence, by choosing a higher cutoff, we effectively remove the denser parts of the wind from the model, thus lowering the \bg \ emission contribution of the wind. The size, on the contrary, increases slightly when the cutoff is higher, since the effective inner diameter of the brightness distribution is enlarged due to the collimation angle of the field lines, see Fig. \ref{fig:WindMagSchem}. The maximum shift magnitude among the photocentre offsets is even more extended for the same reason. \\ \indent
By contrast, for both mass loss rate and the velocity law exponent, parameter increases lead to stronger flux ratios, but smaller emission region characteristic sizes and more compact photocentre shift profiles. For the wind velocity exponent $\beta_{wind}$, this behaviour is straightforward to understand. A higher value for $\beta_{wind}$ decreases the acceleration of the wind. In turn, in requires the wind to travel further to reach the same velocity, which leads to the size increase due to the angle of the stream lines.  The velocity also influences the density inversely, as the density of gas particles in a slow moving outflow is higher, and so is subsequently then the gas emission. The change in characteristic size and photocentre offsets following such a parameter variation is most pronounced at high velocities. As there is now more or less, for decreasing or increasing $\beta_{wind}$ respectively, gas moving at those high velocities, the region appears larger or smaller at those wavelengths. The parallel decrease in flux ratio that comes with a larger region is largely irrelevant to the characteristic size, as long as the hydrogen is dense enough to produce \bg \ emission, since the characteristic size is obtained from the continuum-corrected visibilities. 
\\ \indent
An analogous argument can be made for the changes induced by a variation of the global mass loss rate. The local mass loss rate $\dot{m}$(R) at a certain distance R from the central star is $\propto$ $\dot{M}_{loss}$, so a high global mass loss rate leads to an increase in local mass loss at every distance and thus increased wind density at every distance. However, in this case, the $\alpha_{wind}$ parameter, as well as the chosen collimation angle of the wind stream lines, lead to a stronger increase in density closer to the inner edge of the disk. Under such circumstances, a fit of a Gaussian disk model would yield a smaller characteristic size. \\ \indent 
The stellar parameters appear to again have some of the weakest effects on the three observables. The rotation period in particular does not affect the model at all, since the angular velocity component of the wind is fully determined by Keplerian velocity at the stream line anchor point. As a notable difference compared to the magnetospheric model, the stellar radius seems to rank below both inner radius and width of the wind region, even though both are defined in units of R$_*$ and thus naively the same reasoning as laid out there should apply. The equations governing the density and velocity profiles of the wind model, however, show that dependencies on length scales and sizes mostly come in the form of a ratios between different different projected radial distances, which are all defined in stellar radius units. It follows then that these quantities are relatively less affected by a change in R$_*$. It is also the case here that the most impactful distance scale, the cutoff height z$_{crit}$, is defined in absolute astronomical units and does not vary with the stellar radius. \\ \indent
Finally, we note that the dark disk radius parameter $R_{dd}$, which we introduced to regulate the amount of detected flux from the back side of the disk, is of below average influence according to its computed $\xi_{Tot}$. While this is true when considering the overall profile changes induced by a variation in $R_{dd}$, it does not properly illustrate the effect on profile symmetry. The dark disk radius $R_{dd}$ is the only parameter that allows us to directly adjust the relative strength between blue and red wings of the observable profiles. This unique quality of the parameter is not well reflected in the response metric, given that we consider average metrics over the entire relevant wavelength range.

\onecolumn

\noindent
\clearpage

\section{Azimuth dependency of the non-axisymmetric MA model}
\label{sec:AzimuthAppendix}
\begin{figure*}[h] 
    \centering
    \includegraphics[width=1\linewidth]{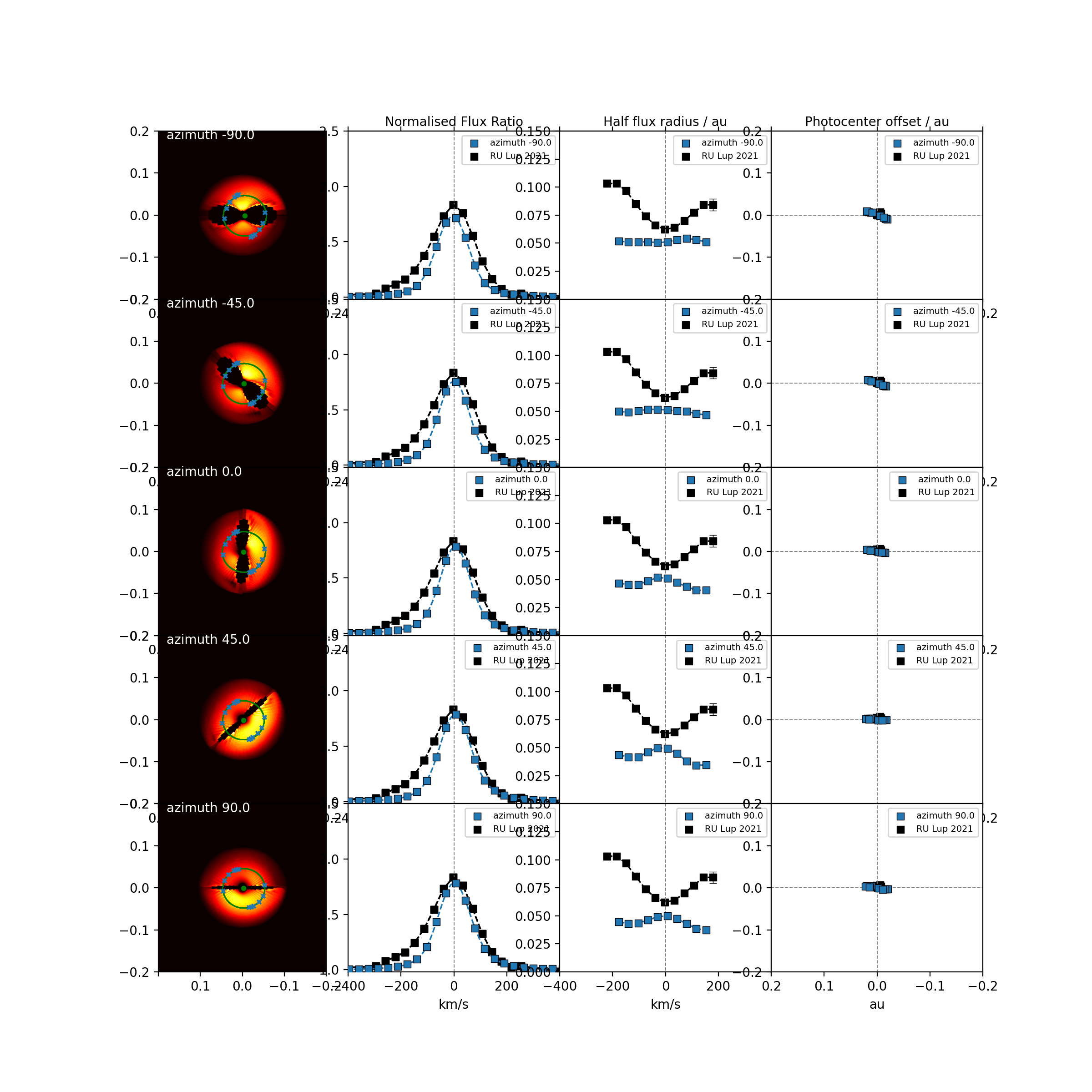}
    \caption{Line-to-continuum flux ratio, characteristic size, and photocentre shift of the MA II model at five different azimuth angles from -90$^{\circ}$ to +90$^{\circ}$. The inclination of the image is fixed at 20$^{\circ}$ to conform to the observational value obtained from VLTI GRAVITY data for the inner disk of RU Lup in 2021. The images depict the emission region at 0 km/s.}
    \label{fig:AzimuthAppendix}
\end{figure*}
\noindent

\section{Grid profiles for HM II}
\label{fig:gridprofiles}
\begin{figure}[h] 
    \centering
    \includegraphics[width=0.34\linewidth]{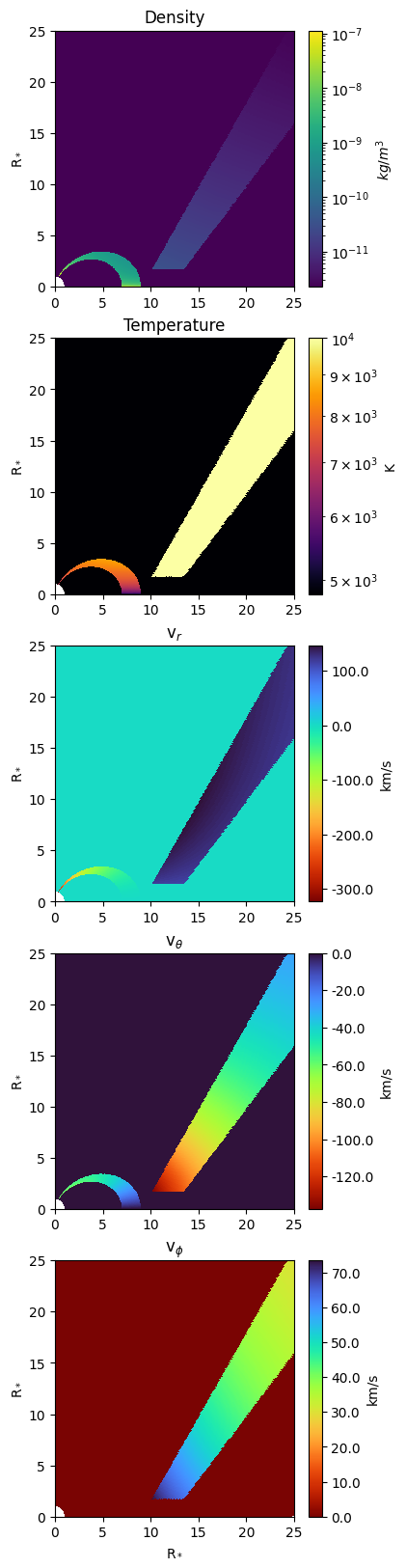}
    \caption{2D model profiles for hybrid model II. Shown are, from top to bottom: Mass density, temperature, and the three spherical components of the velocity field.}\label{fig:dataplot2}
  \end{figure}
  \noindent
\end{document}